\documentclass[aps,prb,twocolumn,showpacs,amsmath]{revtex4}
\usepackage{bm}
\usepackage{graphicx}% Include figure files

\newcommand{\mkright}{draft (\today)} 

\newcommand{\cH}{{\cal H}}
\renewcommand{\Re}{\mbox{Re}\,}
\renewcommand{\Im}{\mbox{Im}\,}

\begin{document}

\title{Flux domes in superconducting films without edges}

\author{John R. Clem}
\affiliation{%
	Ames Laboratory--DOE and Department of Physics and Astronomy, 
	Iowa State University, Ames Iowa 50011, USA}

\author{Yasunori Mawatari}
% \email[]{y.mawatari@aist.go.jp}
\affiliation{%
	National Institute of Advanced Industrial Science and Technology (AIST), 
	Tsukuba, Ibaraki 305--8568, Japan}

\date{\today}
% \received{}\revised{}\accepted{}\published{}

\begin{abstract} Domelike magnetic-flux-density distributions previously have
been observed experimentally and analyzed theoretically in superconducting films
with edges, such as in strips and thin plates.  Such flux domes have
been explained as arising from  a combination of strong geometric
barriers and weak bulk pinning.  In this paper we predict that, even in films
with bulk pinning, flux domes also occur when vortices and antivortices
are produced far from the film edges underneath current-carrying wires,
coils, or permanent magnets placed above the film.  Vortex-antivortex pairs
penetrating through the film are generated when the magnetic field parallel to
the surface exceeds
$H_{c1}+K_c$, where $H_{c1}$ is the lower critical field
and $K_c = j_c d$ is the critical
sheet-current density (the product of the bulk critical current  density
$j_c$ and the film thickness
$d$).  The vortices and antivortices move in
opposite directions to locations where they join others to create
separated vortex and antivortex flux domes.  We consider a simple arrangement of
a pair of  current-carrying wires carrying current $I_0$ in opposite directions
and calculate the magnetic-field and current-density distributions as a function
of
$I_0$ both in the bulk-pinning-free case ($K_c= 0$) and in the presence of
bulk pinning, characterized by a field-independent critical sheet-current
density ($K_c > 0$).
\end{abstract}

% 74.25.Nf	Response to electromagnetic fields
% 74.25.Sv	Critical currents 
% 74.78.-w  Superconducting films and low-dimensional structures 

\maketitle

	\thispagestyle{myheadings}%____________________
	\pagestyle{myheadings}\markright{\mkright}%____________________

\section{Introduction%===========================
\label{Sec_Intro}}

The hysteretic penetration of magnetic flux into a superconductor has long been a
subject of experimental and theoretical interest.  One phenomenon that has often
been observed in flat samples in an increasing perpendicular applied magnetic
field is a domelike magnetic-flux-density distribution centered in the middle of
the sample, surrounded by a superconducting flux-free zone.  This phenomenon has
been  investigated experimentally and interpreted
theoretically in both type-I
superconductors\cite{DeSorbo64,Baird64,Haenssler67,Huebener72,Clem73,Provost74,
Fortini76,
Fortini80,Castro99} and weak-pinning type-II
superconductors.\cite{Indenbom94,Schuster94,Zeldov94,Maksimov95a,Maksimov95b,
Morozov96,
Benkraouda96,Doyle97,Brandt99a,Brandt99b,Mawatari03}
In these previous investigations, the flux domes have been explained as arising
from an energy barrier of geometric origin at the edge of the sample. Once this
barrier is overcome, flux tubes or vortices escape from the edge and are driven to
the middle by Meissner screening currents, which flow on the
sample's surface. 

In this paper we consider a type-II superconducting film without edges and predict
that when it is subjected to local magnetic fields produced by current-carrying
wires  above the sample, vortex and antivortex magnetic-flux domes are produced
in the film when the currents in the wires are sufficiently large.  When bulk
pinning is  weak, we predict that the vortex and antivortex domes are  separated
from  each other, but  as the  bulk pinning increases, the space between the
vortex and antivortex domes shrinks to zero.  

To investigate these effects in an easily calculable geometry, we consider in
Sec.~\ref{Sec_SC-wire} a simple model in which the local magnetic fields are
produced by a pair of infinitely long straight wires.  The resulting
two-dimensional geometry allows us to calculate all the magnetic-field and
sheet-current distributions analytically.  In Sec.~\ref{Sec_Meissner} we 
discuss the Meissner-state response of the film before any penetration of
vortices  into the film, and in Sec.~\ref{Sec_Vortices} we discuss the
distributions produced by vortices and  antivortices that have  penetrated
through the film thickness.  We then present the  magnetic-field and
sheet-current distributions associated with the flux domes, both
in the absence of bulk pinning [Sec.~\ref{Sec_Kc=0}] and in the presence of
weak [Sec.~\ref{Sec_Kc>0}] and strong  [Sec.~\ref{Sec_Kc_strong}] bulk pinning. 
In Sec.~\ref{Sec_conclusion} we summarize our results, discuss the generality
of the predicted effects in more easily realizable experimental configurations,
consider similar phenomena in type-I superconductors, and discuss possible
extensions of this work.  Calculations of screening effects in the Meissner state
are presented in Appendix A, and derivations of the complex field and complex
potential are given in Appendix B.

\section{Superconducting films and linear wires%===========================
\label{Sec_SC-wire}} We consider a simple geometry in which vortex and
antivortex flux domes are produced far from the film edges.  For simplicity we
consider an infinite type-II superconducting film and a pair of
infinitely long current-carrying wires, as shown in Fig.~1.  
\begin{figure}%***** Fig.1 ************************
\includegraphics[width=8cm]{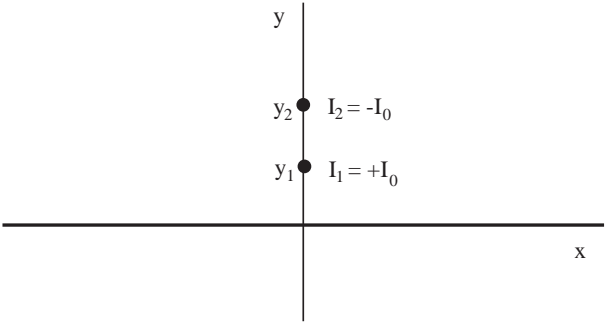}
\caption{%
Wire 1 at $(x,y)=(0,y_1)$ carries current $I_1= I_0$ in the $z$ direction and
wire 2  at $(x,y)=(0,y_2)$ carries current
$I_2 = -I_0$ above an infinite type-II superconducting film in the $xz$ plane.}
\label{fig1}
\end{figure}
A wire, carrying current $I_1=I_0$ parallel to the $z$ axis, is
situated at
$(x,y)=(0,y_1)$, where $y_1>0$.  The return current $I_2 = -I_0$ is carried by a
second wire at $(x,y)=(0,y_2)$, where $y_2>y_1$. The wire radius 
$r_w$ is assumed to be much smaller than either $y_1$ or the wire separation
$y_2-y_1$.  A superconducting film, infinitely extended in the $xz$ plane, is
situated at
$-d/2<y<d/2$, where the film thickness
$d$ is   larger than the London penetration
depth $\lambda$ and $d \ll y_1$.  As discussed in Appendix A, these conditions
guarantee that when the film is in the Meissner state, the magnetic field 
below the film is
negligibly small.

Flux pinning in the
film is characterized by the critical current density
$j_c$, which is assumed to be constant (independent of magnetic field), as in
Bean's critical state model,~\cite{Bean62} and isotropic (independent of vortex
direction)  However, the relevant physical quantity here is the critical
sheet-current density $K_c = j_c d$.  Since we are interested in the case for
which
$d \ll y_1$, in the following we ignore the finite thickness of the film, as
this simplification allows us to obtain  simple analytic expressions for the
magnetic-field and current-density distributions.  

We introduce the complex field $\cH(\zeta)=H_y(x,y)+iH_x(x,y)$, which is
an analytic function of $\zeta=x+iy$ except for poles at $\zeta = \zeta_1 =i
y_1$ and $\zeta = \zeta_2 =
i y_2$ and a branch cut at $y = 0$.  The Biot-Savart law for the complex field
is given by 
\begin{equation}
	\cH(\zeta)= \cH_0(\zeta) +\frac{1}{2\pi} 
		\int_{-\infty}^{+\infty}du \frac{K_z(u)}{\zeta -u} , 
\label{Biot-Savart}
\end{equation} 
where 
\begin{equation}
	\cH_0(\zeta) =\frac{I_0}{2\pi} \frac{1}{\zeta-iy_1} -\frac{I_0}{2\pi}
\frac{1}{\zeta-iy_2} , 
\label{H0}
\end{equation}
is the complex field arising from the pair of
wires alone (see Fig.~1) and $K_z(x)$ is the sheet current in the film.   
The complex potential describing the field generated by the wires alone,
defined by
${\cal G}_0(\zeta)=\int_0^\zeta\cH_0(\zeta')d\zeta'$, is
\begin{equation}
{\cal G}_0(\zeta)=\frac{I_0}{\pi}\ln\Big(\frac{1+i\zeta/y_1}{1+i\zeta/y_2}\Big),
\end{equation}
and the contour lines of the real
part of
${\cal G}_0(\zeta)$ correspond to the magnetic field lines of $\cH_0(\zeta)$. 
At the upper
($\zeta=x+i\epsilon$) and  lower  ($\zeta=x-i\epsilon$) surfaces  of the
superconducting film, (where we take $\epsilon = d/2$ to be a
positive infinitessimal, since $d \ll y_1$,) the perpendicular and parallel
magnetic fields
$H_y(x,0)=\Re\cH(x\pm i\epsilon)$ and $H_x(x,\pm \epsilon)=\Im\cH(x\pm
i\epsilon)$ are obtained from Eq.~\eqref{Biot-Savart} as 
\begin{eqnarray}
	H_y(x,0) &=& H_{0y}(x,0) +\frac{\rm P}{2\pi} 
		\int_{-\infty}^{+\infty}du \frac{K_z(u)}{x-u} , 
\label{Hy_x+-i0}\\
	H_x(x,\pm \epsilon) &=& H_{0x}(x,0) \mp K_z(x)/2 , 
\label{Hx_x+-i0}
\end{eqnarray} 
where $H_{0y}(x,0) = \Re \cH_0(x)$, $H_{0x}(x,0) = \Im \cH_0(x)$, and P denotes
the principal value integral. The complex potential is defined by ${\cal
G}(\zeta)=\int_{i\epsilon}^\zeta \cH(\zeta')d\zeta'$, and the contour lines of the
real part of
${\cal G}(\zeta)$ correspond to the magnetic field lines of $\cH(\zeta)$.

\subsection{Meissner-state response%**************************
\label{Sec_Meissner}} 
We first consider the magnetic-field
distribution when the film is in the (vortex-free) Meissner state and
wires 1 and 2 carry dc currents $I_1=I_0 > 0$ and $I_2 = -I_0 < 0$ after 
monotonically increasing in magnitude from zero.
As discussed in Appendix A, when the current $I_0$ is small and the thickness $d$
is  larger than
$\lambda$, the magnetic field is practically zero below the film,
where
$y=\Im
\zeta <0$.  The field distribution above the film can be obtained by adding to
$\cH_0(\zeta)$ and ${\cal G}_0(\zeta)$ the contributions $\cH_I(\zeta)$ and
${\cal G}_I(\zeta)$ due to image wires at
$\zeta = -\zeta_1 = -iy_1$ and $\zeta = -\zeta_2 = -iy_2$ carrying currents
$-I_0$ and $+I_0$, respectively: 
\begin{eqnarray}
	\cH_I(\zeta)& =&-\frac{I_0}{2\pi} \frac{1}{\zeta+iy_1} +\frac{I_0}{2\pi}
\frac{1}{\zeta+iy_2} , \\
\label{HI}
{\cal
G}_I(\zeta)&=&\frac{I_0}{\pi}\ln\Big(\frac{1-i\zeta/y_2}{1-i\zeta/y_1}\Big).
\label{GI}
\end{eqnarray}
The resulting complex field $\cH_M(\zeta)=\cH_0(\zeta)+\cH_I(\zeta)$ is
\begin{equation}
	\cH_M(\zeta)= 
	\begin{cases} \displaystyle 
	i	\frac{I_0}{\pi}\Big(\frac{y_1}{\zeta^2+y_1^2} -\frac{y_2}{\zeta^2+y_2^2}\Big)
		& \mbox{for } \Im \zeta >0 , \\
		0 & \mbox{for } \Im \zeta <0 . 
	\end{cases}
\label{cH_linear}
\end{equation}
The subscript $M$ is a reminder that this field  describes the
Meissner-state response to the applied field given in Eq.~\eqref{H0}.
Note that 
\begin{equation}
H_{Mx}(x,\epsilon)= 
	\frac{I_0}{\pi}\Big(\frac{y_1}{x^2+y_1^2} -\frac{y_2}{x^2+y_2^2}\Big).
\label{HMx}
\end{equation} 
As is usual using the method of images,
$H_{Mx}(x,\epsilon) = 2 H_{0x}(x,0) =2
\Im
\cH_0(x)$ [see Eq.\ \eqref{H0}].  The corresponding complex potential
${\cal G}_M(\zeta)=\int_{i\epsilon}^\zeta{\cal H}_M(\zeta')d\zeta'$ for $\Im \zeta
>0$ is given by 
\begin{eqnarray}
	{\cal G}_M(\zeta) = 
		i\frac{I_0}{\pi}\Big[\arctan\Big(\frac{\zeta}{y_1}\Big) 
-\arctan\Big(\frac{\zeta}{y_2}\Big)\Big]. 
\label{cG_linear}
\end{eqnarray} 
We may take ${\cal G}_M(\zeta)=0$ for $\Im \zeta <0.$
\begin{figure}%***** Fig.2 ************************
\includegraphics[width=8cm]{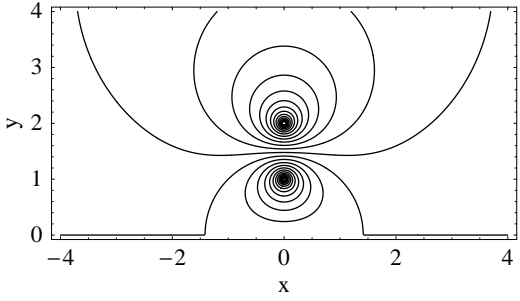}
\caption{%
Contour plot of the real part of the complex potential ${\cal G}_M(x+iy)$ vs
$x$ and $y$ for $y_1 = 1$ and $y_2 = 2$.  The contours correspond to magnetic
field lines of the complex field $\cH_M(x+iy)$ describing the Meissner-state
response of the  superconducting film to currents in the wires shown in Fig.~1.
 The contours near the wires at $\zeta = iy_1$ and $\zeta
=iy_2$ are  nearly circular.}
\label{fig2}
\end{figure}
The perpendicular magnetic field and sheet-current
density are thus given by $H_{My}(x,0)=0$ and 
\begin{equation}
	K_{Mz}(x) = -\frac{I_0}{\pi}\Big( \frac{y_1}{x^2+y_1^2}-
\frac{y_2}{x^2+y_2^2}\Big) . 
\label{KMz_linear}
\end{equation}
The net current induced in the superconducting film is
$\int_{-\infty}^{+\infty}K_{Mz}(x)dx =0$, as expected.  

The maximum magnetic field parallel to the top surface of the film is
\begin{equation}
H_{Mx}(0,\epsilon) = \frac{I_0}{\pi}\Big( \frac{1}{y_1}-
\frac{1}{y_2}\Big) , 
\label{HMx0eps}
\end{equation}
and the maximum magnitude of the sheet-current density is 
\begin{equation}
|K_{Mz}(0)| = \frac{I_0}{\pi}\Big( \frac{1}{y_1}-
\frac{1}{y_2}\Big) . 
\label{KMzmax}
\end{equation}
\begin{figure}%***** Fig.3 ************************
\includegraphics[width=8cm]{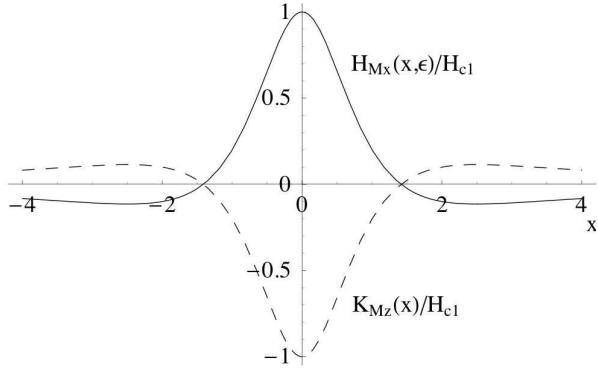}
\caption{%
Parallel magnetic field $H_{Mx}(x,\epsilon)/H_{c1}$ vs $x$ (solid curve) in the
Meissner state at the top surface of the superconducting film when
$I_0=I_{c1}$,  
$H_{Mx}(0,\epsilon)=H_{c1}$, $y_1 = 1$, and $y_2 = 2$.  The dashed curve shows
the corresponding sheet-current density $K_{Mz}(x)/H_{c1}$ vs $x$.  Note that
$K_{Mz}(0) = -H_{c1}.$}
\label{fig3}
\end{figure}

We now define two critical currents as follows.  Let $I_{c1}$ denote the value
of $I_0$ at which $H_{Mx}(0,\epsilon)$ in Eq.\eqref{HMx0eps} reaches the lower
critical field $H_{c1}$,
\begin{equation}
I_{c1}=  \frac{\pi H_{c1}y_1y_2}{y_2-y_1} ,
\label{Ic1}
\end{equation}
and let $I_{c0}$ denote the value of 
$I_0$ at which $|K_{Mz}(0)|$ in Eq.~\eqref{KMzmax} reaches the critical
sheet-current density
$K_c$,
\begin{equation}
I_{c0}=  \frac{\pi K_{c}y_1y_2}{y_2-y_1}.
\label{Ic0}
\end{equation}
Both of these critical currents play important roles in
determining the details of how magnetic flux penetrates through the film.  In the
following sections we discuss in turn the cases for which  
$K_c$ = 0 ($I_{c0}=0$) and  $K_c >
0$ ($I_{c0} > 0$).

\subsection{Vortex-generated fields and currents%**************************
\label{Sec_Vortices}} 

Consider a closely spaced row of vortices along the $z$ axis carrying
magnetic flux $\Phi'$  per unit length up through the
film.  Ignoring spatial variation on the scale of the London penetration
depth $\lambda$ or the intervortex spacing, the magnetic field in the
space $y > 0$ ($y < 0$) appears as if produced by a line of positive
(negative) magnetic monopoles.  At a distance
$r$ from the origin, the magnitude of the magnetic field is $h =
\Phi'/\mu_0 \pi r$.  Expressing this result in terms of a complex magnetic
field $\cH_v(\zeta) = H_{vy}(x,y)+iH_{vx}(x,y)$ and extending it to 
a distribution of vortices or antivortices generating a magnetic field
$H_{vy}(x,0)$ in the plane of the film, we see that the vortex-generated complex
magnetic field can be expressed as
\begin{equation}
\cH_v(\zeta)=\pm \frac{i}{\pi}\int_{-\infty}^{+\infty}du
\frac{H_{vy}(x,0)}{\zeta-u},
\label{Hv}
\end{equation}
where $\zeta = x +iy$ and the upper (lower) sign in Eq.~\eqref{Hv} holds for $y
>0$ ($y<0$). The corresponding vortex-generated sheet-current density
$K_{vz}(x) = H_{vx}(x-i\epsilon)-H_{vx}(x+i\epsilon)$ is
\begin{equation}
K_{vz}(x)=-\frac{\rm P}{\pi}\int_{-\infty}^{+\infty}du
\frac{2H_{vy}(x,0)}{x-u},
\label{Kvz}
\end{equation}
while the Biot-Savart law yields another relation between $K_{vz}(x)$
and $H_{vy}(x,0)$:
\begin{equation}
2H_{vy}(x,0) = \frac{\rm P}{\pi} 
		\int_{-\infty}^{+\infty}du \frac{K_{vz}(u)}{x-u}.
\end{equation}

In the following sections, the complex field can always be
regarded as a linear superposition of the Meissner-state and vortex-generated
complex fields: $\cH(\zeta) = \cH_M(\zeta) + \cH_v(\zeta).$ The complex
potential ${\cal G}(\zeta) = \int_{i\epsilon}^\zeta \cH(\zeta') d\zeta'$ can be
written as
${\cal G}(\zeta) ={\cal
G}_M(\zeta) +{\cal
G}_v(\zeta)$.  Similarly, the sheet-current density can always be expressed as
$K_z(x) = K_{Mz}(x) + K_{vz}(x)$.

\subsection{Flux domes in the absence of bulk pinning%**************************
\label{Sec_Kc=0}} 

In bulk-pinning-free films ($K_c= 0$), the first vortex 
enters the film at $x = 0$, where the maximum magnetic field at the top surface
is equal to the lower critical field
$H_{c1}$.\cite{foot1}  This occurs at the current $I_0 = I_{c1}$, given in
Eq.~\eqref{Ic1} (see Fig.~3).  An  initially tiny vortex loop expands in radius,
and a portion of the loop is driven to the bottom of the film surface, where it
annihilates, resulting in a separated vortex-antivortex pair.  The vortex
(antivortex) carries magnetic flux $\phi_0$ in the $+y$ ($-y$) direction, 
where $\phi_0 = h/2e$ is the
superconducting flux quantum.  Responding to the
Lorentz force
$K_{Mz}(x)
\phi_0$, the
vortex moves in the $x$ direction until it comes to rest at the point $x = x_0 =
\sqrt{y_1y_2}$, where $K_{Mz}(x)$ = 0, as can be seen from
Eq.~\eqref{KMz_linear} and Fig.\ 3.  The antivortex moves in the opposite
direction and comes to rest at the point
$x = -x_0$.  

For increasing values of $I_0$ in the range $I_{c1} < I_0 <  2 I_{c1}$, the
magnetic field distribution perpendicular to the film can be characterized as
having a positive vortex-generated magnetic flux dome in the region
$ b < x < a$, where $0 <b < x_0 < a$, and a  negative
antivortex-generated flux dome in the region $-a < x < -b$.  The complex
magnetic field $\cH(\zeta) = \cH_M(\zeta) +\cH_v(\zeta)$ is given in  
Eq.~\eqref{cH_zero} but with $K_c = 0$.  Subtracting the Meissner-state complex
field
$\cH_M(\zeta)$, we obtain the following expression for the vortex-generated
complex magnetic field, 
\begin{equation}
	\cH_v(\zeta)\!\!=\!\!\frac{I_0}{2\pi} \Big\{ \frac{y_1[\mp i
+\phi(\zeta)/s_1]}{\zeta^2+y_1^2} 
\! -\!\frac{y_2[\mp i
+\phi(\zeta)/s_2]}{\zeta^2+y_2^2} 
		\Big\} ,
\label{Hv0}
\end{equation}
where 
$	s_j= \sqrt{(a^2+y_j^2)(b^2+y_j^2)}$, and the upper (lower) sign holds when
$\zeta = x+iy$ is in the upper (lower) half plane.  Writing
$\phi(\zeta)=\phi_1(x,y)+i\phi_2(x,y)$, we find that
$\phi_1(x,y)$ is  an odd function of $x$ and an even function of $y$ 
[$\phi_1(x,y)=-\phi_1(-x,y)=\phi_1(x,-y)$], while  $\phi_2(x,y)$ is an
even function of $x$ and an odd function of $y$
[$\phi_2(x,y)=\phi_2(-x,y)=-\phi_2(x,-y)$].  Thus $H_{vy}(x,y)$ is an odd
function of $x$ but an even function of $y$, while $H_{vx}(x,y)$ is an odd
function of
$y$ but an even function of $x$.  Just above (below) the real axis,
\begin{equation}
	\phi(x \pm i\epsilon)= 
	\begin{cases} \displaystyle 
	\pm i	\tilde \phi(x),
		& |x|<b  \mbox { or } |x| > a, \\
		\tilde \phi(x), & b<|x|<a, \\
	\end{cases}
\label{phixpm}
\end{equation}
where
\begin{equation}
	\tilde \phi(x)= 
	\begin{cases} \displaystyle 
	\sqrt{(a^2-x^2)(b^2-x^2)},
		& |x|<b , \\
		{\rm sgn}(x) \sqrt{(a^2-x^2)(x^2-b^2)}, & b<|x|<a, \\
-	\sqrt{(x^2-a^2)(x^2-b^2)}, & |x|>a .
	\end{cases}
\label{tildephi}
\end{equation}
From Eq.~\eqref{Hv0} we obtain the following values of $H_{vy}(x,0)$,
$H_{vx}(x,\epsilon)=-H_{vx}(x,-\epsilon),$ and
$K_{vz}(x)=H_{vx}(x,-\epsilon)-H_{vx}(x,\epsilon)$,
\begin{eqnarray}
	H_{vy}(x,0) = 
	\begin{cases} \displaystyle 
 0,  & |x|<b  \mbox { or } |x| > a,\\
\frac{I_0}{2\pi}\Big(\frac{y_1}{s_1(x^2+y_1^2)} \\- 
\frac{y_2}{s_2(x^2+y_2^2)}\Big)\tilde \phi(x), &
 b<|x|<a,
	\end{cases}
\label{Hvyx0_zero}
\end{eqnarray}
\begin{eqnarray}
	H_{vx}(x, \epsilon) = 
	\begin{cases} \displaystyle 
-\frac{I_0}{2\pi}\Big(\frac{y_1[1\!-\! \tilde \phi(x)/s_1]}{x^2+y_1^2} 
- \frac{y_2[1\!-\!\tilde \phi(x)/s_2]}{x^2+y_2^2}\Big),\\ 
\;\;\;\;\;\;\;\;\;\;\;\;\;\; \;\;\;\;\;\;\;\
 |x|<b  \mbox
{ or } |x| > a,\\
 -\frac{I_0}{2\pi}\Big(\frac{y_1}{x^2+y_1^2}
- \frac{y_2}{x^2+y_2^2}\Big), b<|x|<a,
	\end{cases}
\label{Hvxxeps_zero}
\end{eqnarray}
\begin{eqnarray}
	K_{vz}(x)  =  
	\begin{cases} \displaystyle 
\frac{I_0}{\pi}\Big(\frac{y_1[1\!-\! \tilde \phi(x)/s_1]}{x^2+y_1^2} 
- \frac{y_2[1\!-\!\tilde \phi(x)/s_2]}{x^2+y_2^2}\Big),\\ 
\;\;\;\;\;\;\;\;\;\;\;\;\;\; \;\;\;\;\
 |x|<b  \mbox
{ or } |x| > a,\\
 \frac{I_0}{\pi}\Big(\frac{y_1}{x^2+y_1^2}
- \frac{y_2}{x^2+y_2^2}\Big),\;\;\; b<|x|<a,
	\end{cases}
\label{Kzx_zer0}
\end{eqnarray}

\begin{figure}%***** Fig.4 ************************
\includegraphics[width=8cm]{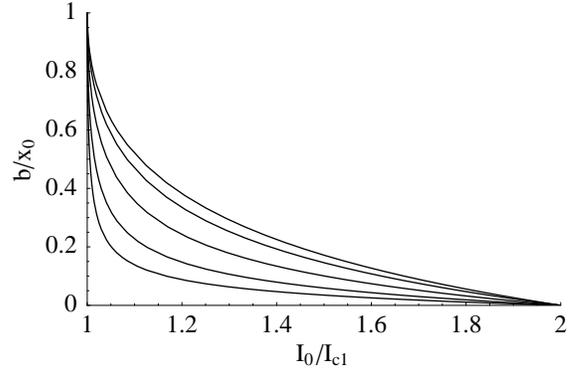}
\caption{%
Plot of the  vortex dome's left boundary $b$ vs $I_0/I_{c1}$ for the
bulk-pinning-free case, where $b$ is in units of $x_0 = \sqrt{y_1y_2}$, for
$I_{c1} < I_0 < 2I_{c1}$ and $y_1/y_2 =$ 0.01, 0.03, 0.1, 0.3, and 1
(bottom to top).
 The right boundary of the vortex dome is at $a = x_0^2/b$.}
\label{fig4}
\end{figure}

As discussed in Appendix B, the requirement that $ \int_{-\infty}^{+\infty}
K_z(x) dx=0 $ leads to the condition that $ab = x_0^2 = y_1 y_2$ when there is
no bulk pinning.  A second condition relating $a$ , $b$ and $I_0$ follows
from the requirement that
$H_x(0,\epsilon)=H_{Mx}(0,\epsilon)+H_{vx}(0,\epsilon)= H_{c1}$, which yields
\begin{equation}
I_0=\Big(\frac{2}{1+y_1y_2(s_1+s_2)/s_1s_2}\Big)I_{c1}.
\label{I0ab}
\end{equation}
Combining these two conditions and eliminating $a$, we obtain the following
connection between $I_0$ and $b$,
\begin{equation}
I_0=\Big(\frac{2}{1+b(y_1+y_2)/\sqrt{(y_1^2+b^2)(y_2^2+b^2)}}\Big)I_{c1}.
\end{equation}
 When $I_0 =I_{c1}$, $a = b = x_0 = \sqrt{y_1
y_2}$.  As
$I_0$ increases above
$I_{c1}$,
$b$ decreases monotonically, as shown in Fig.~\ref{fig4} and the lower solid
curve in Fig.~\ref{fig11}. 
In the limit as $I_0
\to 2 I_{c1}$, we see that $b \to 0$, such that  
$a = y_1 y_2/b 
\to \infty$.  
For $I_0 > 2 I_{c1}$, the magnetic-flux-filled region
extends from $-\infty$ to $+\infty$, $K_z(x) = 0$, and $\cH(\zeta) =
\cH_0(\zeta)$ everywhere, since the superconducting film is then
completely incapable of screening the magnetic field produced by the two wires.

Figures 5 and 6 show plots of $H_y(x,0) = H_{vy}(x,0)$ and  $K_z(x)$ vs
$x$ for several values of
$I_0$ in the range $I_{c1} \le I_0 \le 2I_{c1}$.
\begin{figure}%***** Fig.5 ************************
\includegraphics[width=8cm]{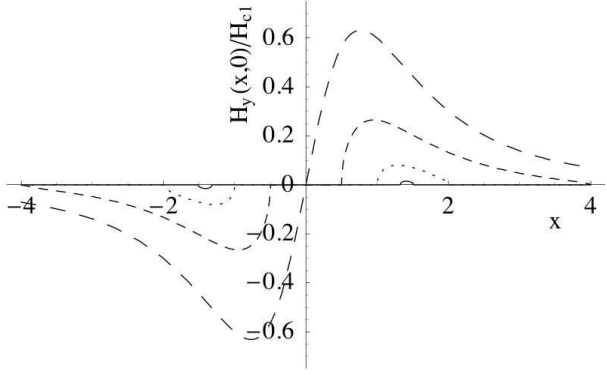}
\caption{%
The perpendicular magnetic field $H_y(x,0)/H_{c1} = H_{vy}(x,0)/H_{c1}$ vs $x$
exhibits vortex and antivortex flux domes  at $b < |x| <a$ in the
absence of bulk pinning.  With
$y_1 =1$,
$y_2 = 2$, and $x_0 = \sqrt{y_1y_2} = 1.414$, when $I_0 = 1.001 I_{c1}$, tiny
domes are centered at $\pm x_0$.  As $I_0$ increases, the domes
 become taller
and wider, as shown for
$I_0 = 1.026 I_{c1}$ when
$b = 1$ (dotted), $I_0 = 1.212 I_{c1}$ when
$b = 0.5$ (short dash), and $I_0 = 2 I_{c1}$ when
$b = 0$ and $a = \infty$ (long dash). For $I_0 \ge 2I_{c1}$, the superconducting
film is no longer capable of screening, $K_z(x) = 0$ everywhere, and $H_y(x,0) =
H_{0y}(x,0)$ [Eq.~\eqref{H0}].}
\label{fig5}
\end{figure} 
\begin{figure}%***** Fig.6 ************************
\includegraphics[width=8cm]{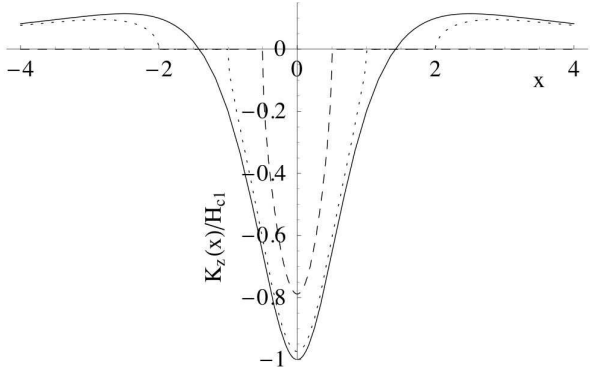}
\caption{%
Sheet-current-density $K_z(x)/H_{c1}$ vs $x$ in the
absence of bulk pinning for
$y_1 =1$,
$y_2 = 2$, and $x_0 = \sqrt{y_1y_2} = 1.414$.  When $I_0 = I_{c1}$, $K_z(x)$
(solid) is equal to $K_{Mz}(x)$, shown in Fig.~3.  With increasing $I_0$,
regions of
$K_z(x) = 0$ develop underneath the vortex and antivortex domes at $b <
|x| <a$, beginning at $x = \pm x_0$.  Shown are results for
$I_0 = 1.026 I_{c1}$ when
$b = 1$ (dotted) and $I_0 = 1.212 I_{c1}$ when
$b = 0.5$ (dashed). For $I_0 \ge 2I_{c1}$, the superconducting
film is no longer capable of screening, and $K_z(x) = 0$ everywhere.}
\label{fig6}
\end{figure} 

Figure 7 shows plots of
$H_x(0,\pm\epsilon)$ = $H_{Mx}(0,\pm\epsilon)+H_{vx}(0,\pm\epsilon)$  vs
$I_0$.
\begin{figure}%***** Fig.7 ************************
\includegraphics[width=8cm]{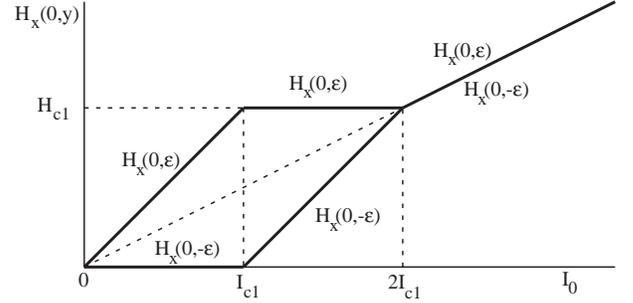}
\caption{%
Parallel magnetic fields at $x=0$ at the top and bottom surfaces
$H_x(0,\epsilon)$ and
$H_x(0,-\epsilon)$ vs
$I_0$ in the absence of bulk pinning.  When $0 \le I_0 \le I_{c1}$,
$H_x(0,\epsilon) = H_{Mx}(0,\epsilon)$ [Eq.~\eqref{cH_linear}] and
$H_x(0,-\epsilon) = H_{Mx}(0,-\epsilon)=0$.  When $ I_{c1} \le I_0 \le
2 I_{c1}$, $H_x(0,\epsilon) =H_{Mx}(0,\epsilon) +H_{vx}(0,\epsilon)= H_{c1}$,
 and
$H_x(0,-\epsilon) = H_{vx}(0,-\epsilon) = -H_{vx}(0,\epsilon)$. When $ I_0 \ge
2 I_{c1}$, $H_x(0,\pm\epsilon) = H_{0x}(0,0) =H_{c1}I_0/2I_{c1}$
[Eq.~\eqref{H0}].}
\label{fig7}
\end{figure} 
\begin{figure}%***** Fig.8 ************************
\includegraphics[width=8cm]{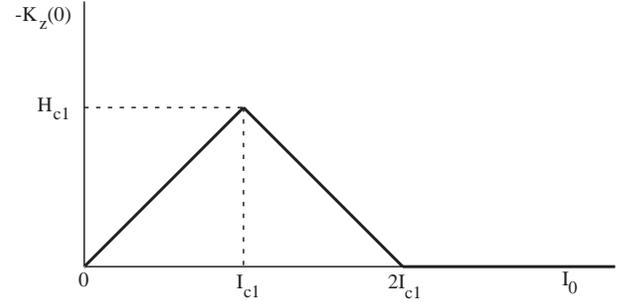}
\caption{%
The magnitude of the sheet-current-density at the origin $-K_z(0) =
H_x(0,\epsilon)-H_x(0,-\epsilon)$ vs $I_0$ in the absence of bulk pinning (see
Fig.~7). }
\label{fig8}
\end{figure} 
For $0 \le I_0 \le I_{c1}$, we have $H_x(0,\epsilon) = H_{Mx}(0,\epsilon) =
H_{c1}I_0/I_{c1}$ and  $H_x(0,-\epsilon) = H_{Mx}(0,-\epsilon) =
0.$  For $I_{c1} \le I_0 \le 2 I_{c1}$, we have $H_x(0,\epsilon) =
H_{Mx}(0,\epsilon) + H_{vx}(0,\epsilon) = H_{c1}$ and $H_x(0,-\epsilon) =
 H_{vx}(0,-\epsilon) = H_{c1}(I_0/I_{c1}-1)$, where
$H_{Mx}(0,\epsilon) = H_{c1}I_0/I_{c1}$ and  $H_{vx}(0,\pm \epsilon) = \pm
H_{c1}(1-I_0/I_{c1}).$ When $I_0 = 2 I_{c1}$, $H_x(0,\pm
\epsilon) = H_{c1}$.  For $I_0 > I_{c1}$, $H_x(0,\pm \epsilon) = H_{0x}(0,0) =
H_{c1}I_0/2I_{c1}$,  and in addition $H_x(x,\pm \epsilon) = H_{Mx}(x, \pm
\epsilon) + H_{vx}(x, \pm \epsilon) = H_{0x}(x,0)$ for all $x$.  The macroscopic
magnetic-field distribution is then the same as it would be if the
superconducting film were absent.

Figure 8 shows a plot of 
$-K_z(0)$ vs
$I_0$, where  $K_z(0) =
H_x(0,-\epsilon)-H_x(0,\epsilon)$.  For $0 \le I_0 \le I_{c1}$,
$-K_z(0) = H_{c1}I_0/I_{c1}$, and for $I_{c1} \le I_0 \le 2 I_{c1}$,
$-K_z(0) = H_{c1}(2-I_0/I_{c1})$.  For $I_0 > 2I_{c1}$, $K_z(x) = 0$ for
all $x$, i.e., everywhere in the film.

The vortex-generated complex potential ${\cal G}_v(\zeta)={\cal
G}(\zeta)-{\cal G}_M(\zeta)$ [see Eqs.~\eqref{cG_linear} and \eqref{cG_zero}]
in the absence of bulk pinning is  
\begin{eqnarray}
	{\cal G}_v(\zeta)\!\! &=\!\!& \frac{I_0}{2\pi} 
		\left[ g_v(\zeta,y_1) -g_v(\zeta,y_2) \right] , 
\label{cG_v}
\end{eqnarray}
where 
\begin{eqnarray}
	g_v(\zeta,y) 
	&=& \mp i\arctan (\zeta/y) \nonumber\\
	&& {}\mp\frac{i}{asy} [ a^2y^2\bm{E}(\theta,k) 
		+y^2(b^2+y^2)\bm{F}(\theta,k)\nonumber\\
	&& -(a^2+y^2)(b^2+y^2)\bm{\Pi} 
		(\theta,-b^2/y^2,k) ], 
\label{g_v} \\
	s &=& \sqrt{(a^2+y^2)(b^2+y^2)}, \\
	\theta &=& \arcsin(\zeta/b), \mbox{ and } \\
	k &=& b/a , 
\end{eqnarray}
where $\bm E$, $\bm K$, and $\bm \Pi$ are incomplete elliptic integrals and the
upper (lower) signs hold in the upper (lower) half $\zeta$ plane. 
\begin{figure}%***** Fig.9 ************************
\includegraphics[width=8cm]{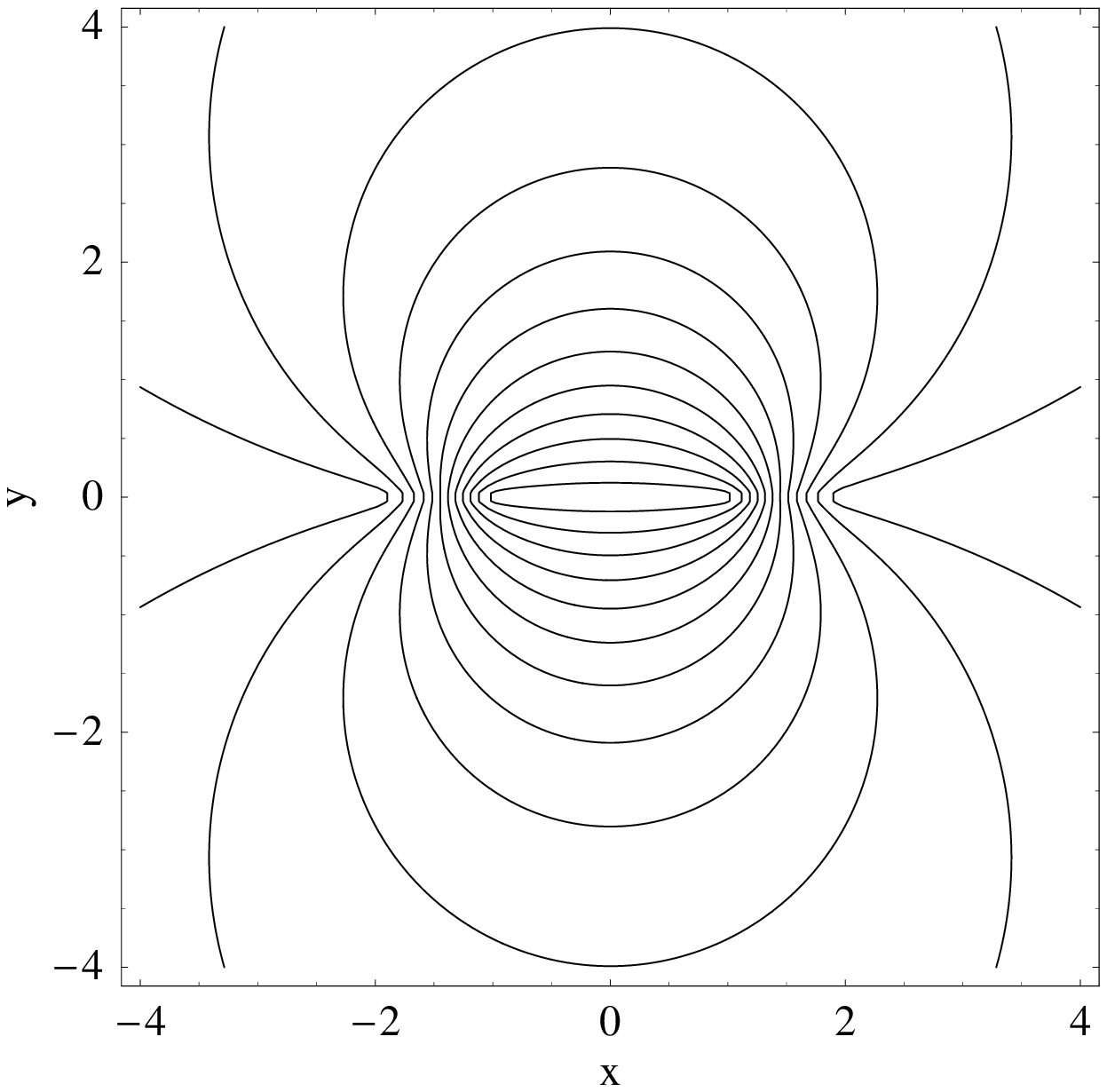}
\caption{%
Contour plot of the real part of the vortex-generated complex potential in the
absence of bulk pinning [Eq.~\eqref{cG_v}]
${\cal G}_v(x+iy)$ vs
$x$ and $y$ for $y_1 = 1$, $y_2 = 2$, $a = 2$, and $b = 1$, as for the dotted
curves in Figs.~5 and 6.  The contours correspond to magnetic field lines of the
vortex-generated complex field
$\cH_v(x+iy)$ given in Eq.~\eqref{Hv0}. }
\label{fig9}
\end{figure} 
Shown
in Fig.~9 is a contour plot of the real part of  ${\cal G}_v(x+iy)$. 
These contours correspond to the magnetic field lines of the
vortex-generated magnetic field.  The magnetic field flows in a
generally counterclockwise direction, carried by vortices  in the region $b < x <
a$ up through the  film and by antivortices in the region $-a < x < -b$ back down
through the film.  A contour plot of the real part of  ${\cal G}(x+iy)=
{\cal G}_M(x+iy)+{\cal G}_v(x+iy)$ would also show the magnetic field lines
generated by the two wires, as in Fig.~2.

\subsection{Flux domes in the presence of weak bulk
pinning%**************************
\label{Sec_Kc>0}} 

In superconducting films in which bulk pinning is present and is characterized
by a field-independent critical sheet-current density 
$K_c  = j_c d >0$, the first vortex enters the film at $x = 0$, when  the
maximum  parallel magnetic field  at the top surface $H_{Mx}(0,\epsilon)$ [Eq.
\eqref{HMx0eps}] is equal to the lower critical field
$H_{c1}$.\cite{foot1}  As in Sec.~\ref{Sec_Kc=0}, this again occurs at the current
$I_0 = I_{c1}$, given in Eq.~\eqref{Ic1}.  According to
critical-state theory,\cite{Campbell72,Clem79} this vortex advances toward the
bottom surface at the leading edge of a curving flux front of nearly parallel
vortices of thickness $d_p =  [H_{Mx}(0,\epsilon)-H_{c1}]/j_c$, and it reaches
the bottom surface when $d_p = d$, i.e., when
$H_{Mx}(0,\epsilon) = H_{c1} + K_c$ or $I_0 = I_{c1} + I_{c0}$ [see Eqs. \eqref{HMx0eps}, \eqref{Ic1},
and \eqref{Ic0}].  The positive end of the first vortex
is then at $x = x_p$ and its negative end is at $x = -x_p$, where $x_p$ is the
solution of $H_{Mx}(x_p,\epsilon) = H_{c1}$ [see Eq. \eqref{HMx}].  It
can be shown that $x_p \gg d$ when $d \ll y_1$ except when $K_c \ll H_{c1}$.

Once the first vortex reaches the bottom surface, the portion at $x \approx 0$
annihilates with its image, and the vortex divides into two halves.  The half
in the region $x > 0$, which we call a vortex, carries magnetic flux $\phi_0$
{\it up} from the bottom to the top surface, and the  half
in the region $x < 0$, which we call an antivortex, carries magnetic flux
$\phi_0$ {\it down} from the top to the bottom surface.  Since
$H_{Mx}(0,-\epsilon)=0$  and $H_{Mx}(0,\epsilon) = H_{c1} + K_c$ at $I_0=I_{c1}
+ I_{c0}$, the sheet-current density is initially $K_{Mz}(0) =
-H_{c1} - K_c$.  Because $|K_{Mz}(0)| > K_c$, the vortex separates from the rest
of the flux front and is driven in the $x$ direction by the Lorentz force
$K_{Mz}(x)\phi_0$ until it comes to rest
at $x = x_c$, where $|K_{Mz}(x_c)|=K_c$ [see Eq. \eqref{KMz_linear}] and   the
Lorentz force is balanced by the pinning force.
 Similarly, the antivortex moves in the
opposite direction and comes to rest at the point $x = -x_c$.
As we will show below, $0 < x_c < x_0 =
\sqrt{y_1y_2}$ when $ K_c>0$ [see Fig.\ 10]. 

For increasing values of $I_0$ in the range $I_{c1}+I_{c0} < I_0 <
2I_{c1}+ I_{c0}$,
the magnetic field distribution perpendicular to the film can be characterized
as having a positive vortex-generated magnetic flux dome in the region
$ b < x < a$, where $0 <b < x_c < a$, and a  negative
antivortex-generated flux dome in the region $-a < x < -b$.  The complex
magnetic field $\cH(\zeta) = \cH_M(\zeta) +\cH_v(\zeta)$ is given by
Eq.~\eqref{cH_zero}.  Subtracting the Meissner-state complex field
$\cH_M(\zeta)$, we obtain the following expression for the vortex-generated
complex magnetic field, 
\begin{eqnarray}
	\cH_v(\zeta)\!\!&=\!\!&\frac{I_0}{2\pi} \Big\{ \frac{y_1[\mp i
+\phi(\zeta)/s_1]}{\zeta^2+y_1^2} 
\! -\!\frac{y_2[\mp i
+\phi(\zeta)/s_2]}{\zeta^2+y_2^2} 
		\Big\}\nonumber \\
&&\pm i K_c/2,
\label{Hvweak} 
\end{eqnarray}
where 
$	s_j= \sqrt{(a^2+y_j^2)(b^2+y_j^2)}$, and the upper (lower) sign holds when
$\zeta = x+iy$ is in the upper (lower) half plane.  Following a procedure
similar to that used in Sec.~\ref{Sec_Kc=0}, we obtain the following values of
$H_{vy}(x,0)$,
$H_{vx}(x,\epsilon)=-H_{vx}(x,-\epsilon),$ and
$K_{vz}(x)=H_{vx}(x,-\epsilon)-H_{vx}(x,\epsilon)$,
\begin{eqnarray}
	H_{vy}(x,0) = 
	\begin{cases} \displaystyle 
 0,  & |x|<b  \mbox { or } |x| > a,\\
\frac{I_0}{2\pi}\Big(\frac{y_1}{s_1(x^2+y_1^2)} \\- 
\frac{y_2}{s_2(x^2+y_2^2)}\Big)\tilde \phi(x), &
 b<|x|<a,
	\end{cases}
\label{Hvyx0_small}
\end{eqnarray}
\begin{eqnarray}
	H_{vx}(x, \epsilon) = 
	\begin{cases} \displaystyle 
-\frac{I_0}{2\pi}\Big(\frac{y_1[1\!-\! \tilde \phi(x)/s_1]}{x^2+y_1^2} 
- \frac{y_2[1\!-\!\tilde \phi(x)/s_2]}{x^2+y_2^2}\Big)\\ 
\;\; +K_c/2, \;\;\;\;\;\;\;
 |x|<b  \mbox
{ or } |x| > a,\\
 -\frac{I_0}{2\pi}\Big(\frac{y_1}{x^2+y_1^2}
- \frac{y_2}{x^2+y_2^2}\Big) +K_c/2,\\ 
\;\;\;\;\;\;\;\;\;\;\;\;\;\; \;\;\;\;\;\;\;\;\; b<|x|<a,
	\end{cases}
\label{Hvxxeps_small}
\end{eqnarray}
\begin{eqnarray}
	K_{vz}(x)  =  
	\begin{cases} \displaystyle 
\frac{I_0}{\pi}\Big(\frac{y_1[1\!-\! \tilde \phi(x)/s_1]}{x^2+y_1^2} 
- \frac{y_2[1\!-\!\tilde \phi(x)/s_2]}{x^2+y_2^2}\Big)\\\;\;-K_c, 
\;\;\;\;\;\;\;\;\;\;\;\;\;\; \;\;\;\;
 |x|<b  \mbox
{ or } |x| > a,\\
 \frac{I_0}{\pi}\Big(\frac{y_1}{x^2+y_1^2}
- \frac{y_2}{x^2+y_2^2}\Big)-K_c,\; b<|x|<a,
	\end{cases}
\label{Kzx_small}
\end{eqnarray}

\begin{figure}%***** Fig.10 ************************
\includegraphics[width=8cm]{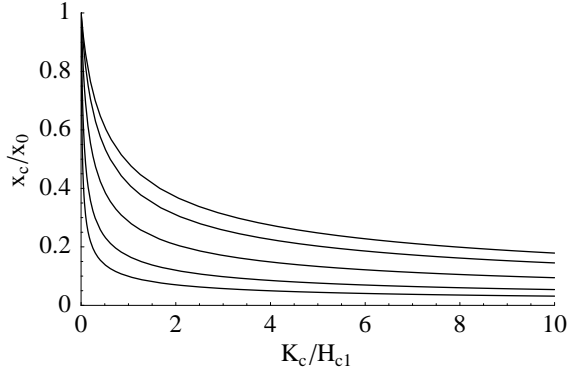}
\caption{%
Plot of $\tilde x_c = x_c/x_0$ vs $K_c/H_{c1}$  in the presence of bulk
pinning, where $x_c$ is the position of the first entering vortex when
$I_0$ just exceeds 
$I_{c1}+I_{c0}$ and 
$x_0 =
\sqrt{y_1y_2}$. 
Curves are shown for
$y_1/y_2 =$ 0.01, 0.03, 0.1, 0.3, and 1 (bottom to top). }
\label{fig10}
\end{figure} 

For $I_{c1}+I_{c0} < I_0 < 2I_{c1}+I_{c0}$, the requirement that
$\int_{-\infty}^{+\infty}K_z(x) dx = 0$ leads to Eq.~\eqref{condition2}.  A
second condition on $a$, $b$, and $I_0$ follows from the requirement that
$H_x(0,\epsilon) = H_{c1}+K_c$, which from Eq.~\eqref{cH_zero} yields: 
\begin{eqnarray}
	\cH_x(0,\epsilon)&=&\frac{I_0}{2\pi} \Big[
\frac{(1+ab/s_1)}{y_1} 
\! -\!\frac{(1+ab/s_2)}{y_2} 
		\Big]+\frac{K_c}{2}\nonumber \\
&=& H_{c1}+K_c.
\label{Hweakcondition} 
\end{eqnarray}
Elimination of $I_0$ between Eqs.~\eqref{condition2} and \eqref{Hweakcondition}
yields the equation
\begin{equation}
\frac{y_1y_2(y_1s_2-y_2s_1)}{(y_2-y_1)s_1s_2+ab(y_2s_2-y_1s_1)}=
\frac{K_c/2H_{c1}}{1+K_c/2H_{c1}}.
\label{abKcweak}
\end{equation}
Numerical solutions of Eqs.~\eqref{Hweakcondition} and \eqref{abKcweak} yield
$a$ and $b$ as a function of $I_0$ for any given value of $K_c$.  As discussed
above, when $I_0$ just exceeds
$I_{c1}+I_{c0}$, the first vortex (antivortex) comes to rest at $x_c$
(-$x_c$).  The equation determining the value of $\tilde x_c = x_c /x_0 = x_c
/\sqrt{y_1y_2}$ can be obtained from Eq.~\eqref{condition2} by setting
$a = b = x_c$ and making use of Eqs.~\eqref{Ic1} and \eqref{Ic0}:
\begin{equation}
\frac{(1-\tilde x_c^2)}{(\tilde x_c^2+y_1/y_2)(\tilde x_c^2+y_2/y_1)}
=\frac{K_c/H_{c1}}{1+K_c/H_{c1}}.
\label{xc}
\end{equation}
Figure 10 shows plots of $\tilde x_c$ vs $K_c/H_{c1}$  for various values of
$y_1/y_2$, obtained by numerically solving Eq.~\eqref{xc}.  Note that for
each case
$\tilde x_c = 1$ when
$K_c$ = 0, and 
$\tilde x_c \to 0$ when $K_c/H_{c1} \to \infty$.  For $K_c/H_{c1}\gg 1$,
\begin{equation}
\tilde x_c \approx 1/\sqrt{(1+y_1/y_2+y_2/y_1)(K_c/H_{c1})}.
\label{tildexclimit}
\end{equation}

Numerical solutions for $a$ and $b$ when $I_{c1}+I_{c0} < I_0 < 2I_{c1}+I_{c0} $
are shown by the dotted curves in Fig.~11 for the case that 
$y_1 = 1$,
$y_2 = 2$, $x_0 = \sqrt{2}$, and
$K_c = H_{c1}/2$, such that $I_{c0} = I_{c1}/2$. 
Note that
$a = b = x_c = 0.585 x_0 = 0.827$ at $I_0 = I_{c1}+I_{c0}$ when vortex
penetration first occurs.
From Eqs. \eqref{Hweakcondition}, \eqref{Ic1}, and \eqref{Ic0}, we see that
$b = 0$ when
$I_0 = 2I_{c1}+I_{c0}.$  When $I_0 > 2I_{c1}+I_{c0}$, $b$ remains equal to
zero,  and the value of
$a$ must be obtained from Eq.~\eqref{condition3}.  Figure 11 also shows similar
plots of  $a$ and
$b$ for $K_c = H_{c1}$ and $K_c = 2 H_{c1}$

\begin{figure}%***** Fig.11 ************************
\includegraphics[width=8cm]{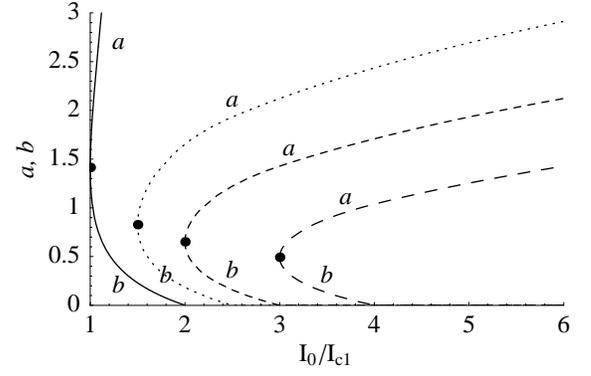}
\caption{%
Plots of $a$ (upper curves), the right boundary of the vortex dome, and $b$
(lower curves), the left boundary, vs
$I_0/I_{c1}$ for  
$y_1 = 1$ and 
$y_2 = 2$:  $K_c=0$ (solid curves separated by the dot at $ a= b=x_c
= 1.414$), $K_c = H_{c1}/2$   (dotted  curves and dot at
$x_c = 0.827$),
$K_c = H_{c1}$ (short-dashed curves  and dot at $x_c =
0.651$), and  $K_c = 2H_{c1}$ (long-dashed curves and  dot at $x_c =
0.493$). Note that $b=0$ when $I_0/I_{c1} \ge 2 +
K_c/H_{c1}$.}
\label{fig11}
\end{figure} 

Figures 12 and 13 show plots of $H_y(x,0) = H_{vy}(x,0)$ and  $K_z(x)$ vs
$x$ for several values of
$I_0$ in the range $I_{c1} +I_{c0} < I_0 \le 2I_{c1} +I_{c0}$.
\begin{figure}%***** Fig.12 ************************
\includegraphics[width=8cm]{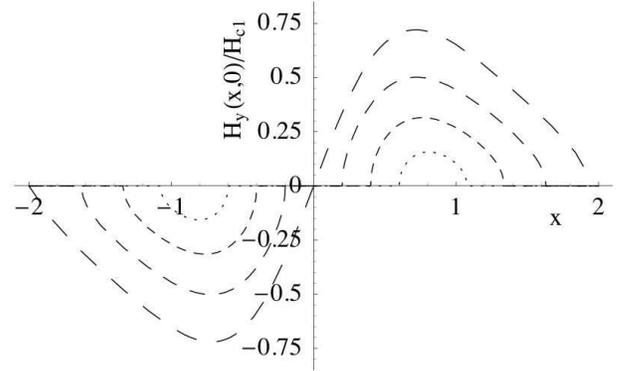}
\caption{%
When  $I_0 > I_{c1}+I_{c0}$, the perpendicular magnetic field
$H_y(x,0)/H_{c1} = H_{vy}(x,0)/H_{c1}$ vs
$x$ initially exhibits separated vortex and antivortex flux domes  at $b < |x|
<a$ even in the presence of bulk pinning, as shown here for $K_c=H_{c1}/2$,
$y_1 =1$,
$y_2 = 2$, and $x_c = 0.827$. When $I_0$ is just above $I_{c1}+I_{c0} = 1.5
I_{c1}$, tiny domes are centered at $\pm x_c$.  As $I_0$ increases, the domes
 become taller
and wider, as shown for
$I_0 = 1.545 I_{c1}$ when $b = 0.600$ and $a = 1.077$ (dotted curves), 
$I_0 = 1.683 I_{c1}$ when $b = 0.400$ and $a = 1.337$ (short-dashed curves), 
$I_0 = 1.966 I_{c1}$ when $b = 0.200$ and $a = 1.630$ (medium-dashed curves),
and $I_0 = 2 I_{c1} + I_{c0} = 2.5 I_{c1}$ when
$b = 0$ and $a = 1.933$ (long-dashed curves).}
\label{fig12}
\end{figure} 
\begin{figure}%***** Fig.13 ************************
\includegraphics[width=8cm]{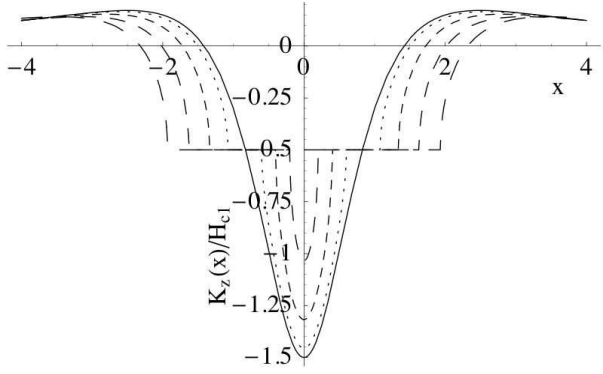}
\caption{%
Sheet-current density $K_z(x,0)/H_{c1}$ vs $x$ in the
presence of
bulk pinning,  shown here for $K_c=H_{c1}/2$,
$y_1 =1$,
$y_2 = 2$, and $x_c = 0.827$.  When $I_0 = I_{c1}+I_{c0}$, $K_z(x)$
(solid) is equal to $K_{Mz}(x)$, shown in Fig.~3.  With increasing $I_0$,
regions of
$K_z(x) = -K_c$ develop underneath the vortex and antivortex domes at $b <
|x| <a$, beginning at $x = \pm x_c$, while  $|K_z(x)| >  K_c$ in the
vortex-free  region $-b < x < b$.  Shown are results for
$I_0 = 1.545 I_{c1}$ when $b = 0.600$ and $a = 1.077$ (dotted curves), 
$I_0 = 1.683 I_{c1}$ when $b = 0.400$ and $a = 1.337$ (short-dashed curves), 
$I_0 = 1.966 I_{c1}$ when $b = 0.200$ and $a = 1.630$ (medium-dashed curves),
and $I_0 = 2 I_{c1} + I_{c0} = 2.500 I_{c1}$ when
$b = 0$ and $a = 1.933$ (long-dashed curves). 
For $I_0 \ge 2I_{c1}+I_{c0}$, $K_z(x) =
-K_c$ in the region $-a < x < a$ and $|K_z(x)| < K_c$ outside this region.}
\label{fig13}
\end{figure} 

Figure 14 shows plots of
$H_x(0,\pm\epsilon)$ = $H_{Mx}(0,\pm\epsilon)+H_{vx}(0,\pm\epsilon)$ vs
$I_0$.
\begin{figure}%***** Fig.14 ************************
\includegraphics[width=8cm]{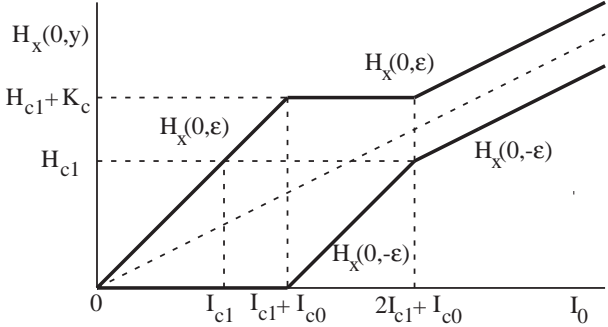}
\caption{%
Parallel magnetic fields at $x=0$ at the top and bottom
surfaces $H_x(0,\epsilon)$ and $H_x(0,-\epsilon)$ vs $I_0$ in the presence
of bulk pinning,  shown here for $K_c=H_{c1}/2$, such that $I_{c1}+I_{c0}= 1.5
I_{c1}$ and $2I_{c1}+I_{c0}= 2.5 I_{c1}$.  When
$0
\le I_0
\le I_{c1}+I_{c0}$,
$H_x(0,\epsilon) = H_{Mx}(0,\epsilon)$ [Eq.~\eqref{HMx}] and
$H_x(0,-\epsilon) = H_{Mx}(0,-\epsilon)=0$.  When $ I_{c1}+I_{c0} \le I_0 \le
2I_{c1}+I_{c0}$, $H_x(0,\epsilon) =H_{Mx}(0,\epsilon) +H_{vx}(0,\epsilon)=
H_{c1}+K_c$,
 and
$H_x(0,-\epsilon) = H_{vx}(0,-\epsilon) = -H_{vx}(0,\epsilon)$. When $ I_0 \ge
2I_{c1}+I_{c0}$, $H_x(0,\pm\epsilon) = H_{0x}(0,0)\pm K_c/2$.}
\label{fig14}
\end{figure} 
\begin{figure}%***** Fig.15 ************************
\includegraphics[width=8cm]{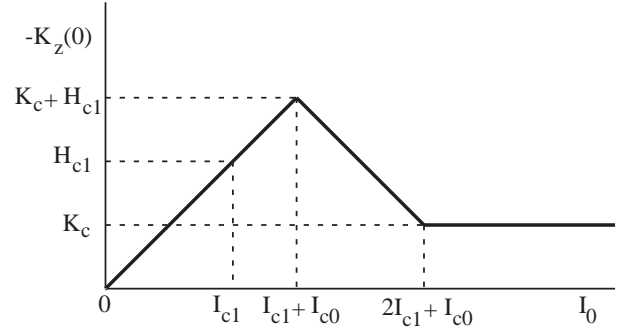}
\caption{%
Magnitude of the sheet-current density at the origin $-K_z(0) =
H_x(0,\epsilon)-H_x(0,-\epsilon)$ vs
$I_0$ in the presence of weak
bulk pinning,  shown here for $K_c=H_{c1}/2$, such that $I_{c2} =2I_{c1}-I_{c0}
= 1.5 I_{c1}$
 (see Fig.~14). }
\label{fig15}
\end{figure}
For $0 \le I_0 \le I_{c1}+I_{c0}$, we have $H_x(0,\epsilon) = H_{Mx}(0,\epsilon)
= H_{c1}I_0/I_{c1}$ and  $H_x(0,-\epsilon) = H_{Mx}(0,-\epsilon) =
0.$  For $I_{c1}+I_{c0} \le I_0 \le 2I_{c1}+I_{c0}$, we have $H_x(0,\epsilon) =
H_{Mx}(0,\epsilon) + H_{vx}(0,\epsilon) = H_{c1}+K_c$ and $H_x(0,-\epsilon) =
 H_{vx}(0,-\epsilon)$, where
$H_{Mx}(0,\epsilon) = H_{c1}I_0/I_{c1}$ and  $H_{vx}(0,\pm \epsilon) = \mp
H_{c1}(I_0-I_{c1}-I_{c0})/I_{c1}.$   When $I_0 =  2I_{c1}+I_{c0}$,
$H_x(0,\epsilon) = H_{c1}+K_c$ and
$H_x(0,-\epsilon) = H_{c1}$.  For
$I_0 \ge  2I_{c1}+I_{c0}$, $H_x(0,\pm \epsilon) = H_{0x}(0,0) \pm K_c/2 =
H_{c1}I_0/2I_{c1} \pm K_c/2$.

Figure 15 shows a plot of
$-K_z(0)$ vs
$I_0$, where  $K_z(0) =
H_x(0,-\epsilon)-H_x(0,\epsilon)$.  For $0 \le I_0 \le I_{c1}+I_{c0}$,
$-K_z(0) = (H_{c1}+K_c)I_0/(I_{c1}+I_{c0})$, and for $I_{c1}+I_{c0} \le I_0 \le
2I_{c1}+I_{c0}$,
$-K_z(0) = K_c +H_{c1}(2I_{c1}+I_{c0}-I_{0})/I_{c1}$.  For $I_0 \ge
2I_{c1}+I_{c0}$,
$-K_z(0) = K_c$.

The vortex-generated complex potential ${\cal G}_v(\zeta)={\cal
G}(\zeta)-{\cal G}_M(\zeta)$ [see Eqs.~\eqref{cG_linear} and \eqref{cG_zero}]
in the presence of weak bulk pinning ($0 < K_c < H_{c1}$) is, when
$I_{c1} < I_0 < I_{c2}$, such that $0 < b < a$,  
\begin{eqnarray}
	{\cal G}_v(\zeta)\!\! &=\!\!& \frac{I_0}{2\pi} 
		\left[ g_v(\zeta,y_1) -g_v(\zeta,y_2) \right] 
 \pm i\frac{K_c}{2}\zeta , 
\label{cG_vweak}
\end{eqnarray}
where $g_v(\zeta,y)$ is given in Eq.~\eqref{g_v}.
\begin{figure}%***** Fig.16 ************************
\includegraphics[width=8cm]{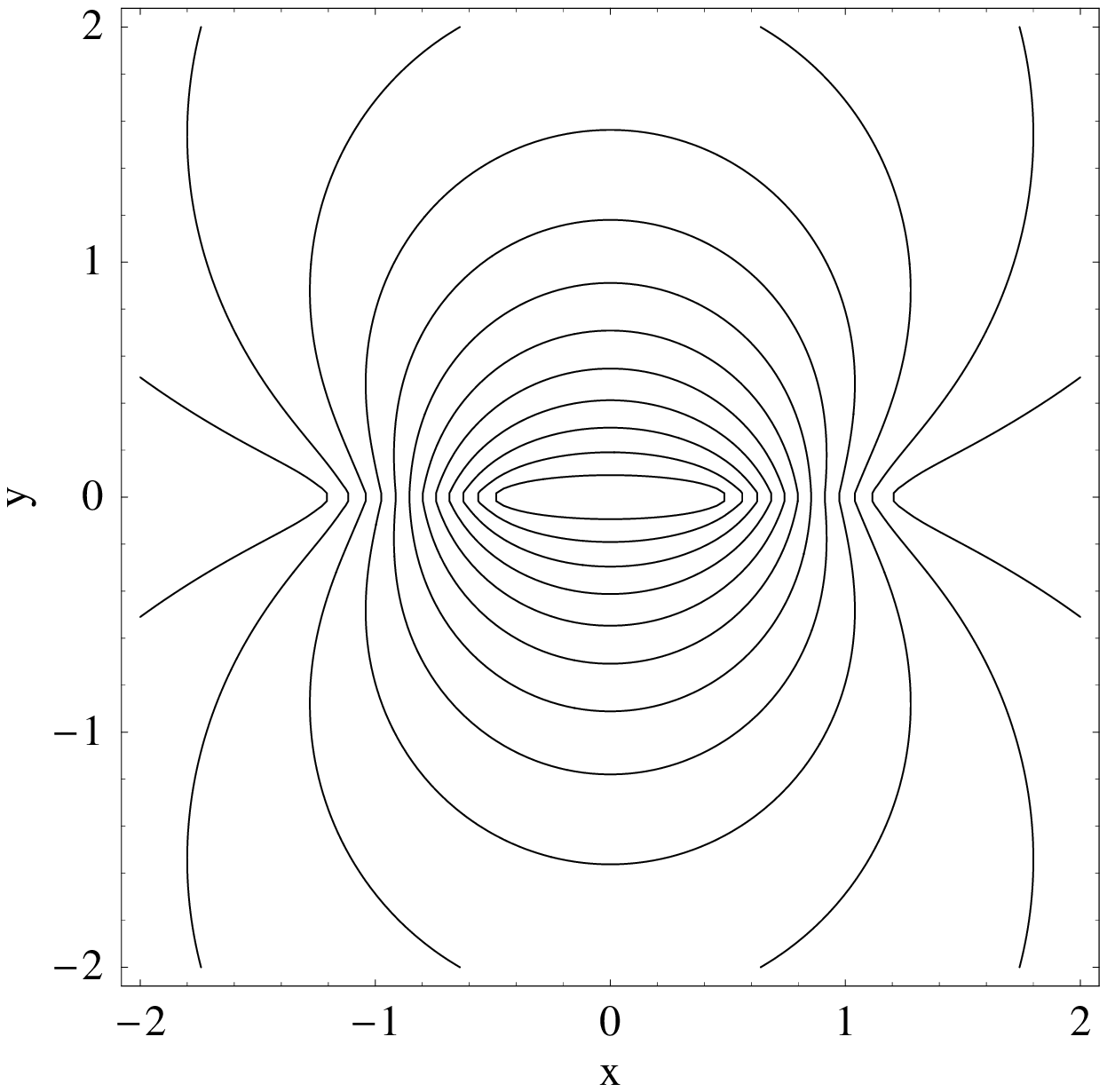}
\caption{%
Contour plot of the real part of the vortex-generated complex potential in the
presence of weak bulk pinning [Eq.~\eqref{cG_vweak}]
${\cal G}_v(x+iy)$ vs
$x$ and $y$ for $y_1 = 1$, $y_2 = 2$, $K_c = H_{c1}/2,$ $I_0 = 1.683  I_{c1}$,
$a = 1.337$,  and
$b = 0.4$, as for the short-dashed curves in Figs.~12 and 13.  The contours
correspond to magnetic field lines of the vortex-generated complex field
$\cH_v(x+iy)$ given in Eq.~\eqref{Hvweak}. }
\label{fig16}
\end{figure} 
Shown
in Fig.~16 is a contour plot of the real part of  ${\cal G}_v(x+iy)$. 
These contours correspond to the magnetic field lines of the
vortex-generated magnetic field.  The magnetic field flows in a
generally counterclockwise direction, carried by vortices  in the region $b < x <
a$ up through the  film and by antivortices in the region $-a < x < -b$ back down
through the film.  A contour plot of the real part of  ${\cal G}(x+iy)=
{\cal G}_M(x+iy)+{\cal G}_v(x+iy)$ would also show the magnetic field lines
generated by the two wires, as in Fig.~2.

\begin{figure}%***** Fig.17 ************************
\includegraphics[width=8cm]{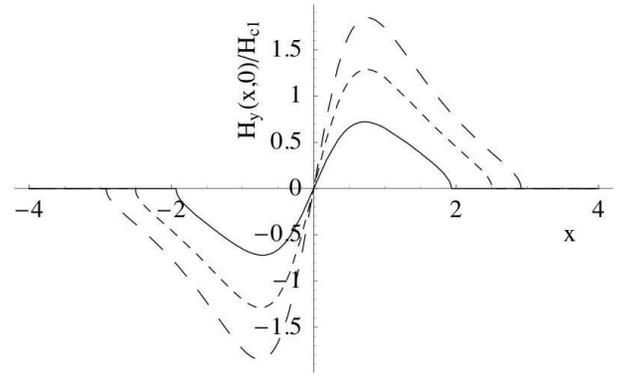}
\caption{%
When $I_0 \ge 2I_{c1} +I_{c0}$ in the presence of 
bulk pinning, the perpendicular magnetic field $H_y(x,0)/H_{c1}
= H_{vy}(x,0)/H_{c1}$ vs
$x$ exhibits adjacent vortex and antivortex flux domes  at $0 < |x| <a$,  as
shown here for
$K_c=H_{c1}/2$, $I_{c0} = 0.5 I_{c1}$,
$y_1 =1$, and
$y_2 = 2$.  As
$I_0$ increases, the domes
 become taller
and wider, as shown for
$I_0 = 2.5 I_{c1}$ when
$a = 1.933$ (solid curves), 
$I_0 = 4.25 I_{c1}$ when $a = 2.504$ (short-dashed curves), and $I_0 = 6 I_{c1}$
when
$a = 2.913$ (long-dashed curves).}
\label{fig17}
\end{figure}

\begin{figure}%***** Fig.18 ************************
\includegraphics[width=8cm]{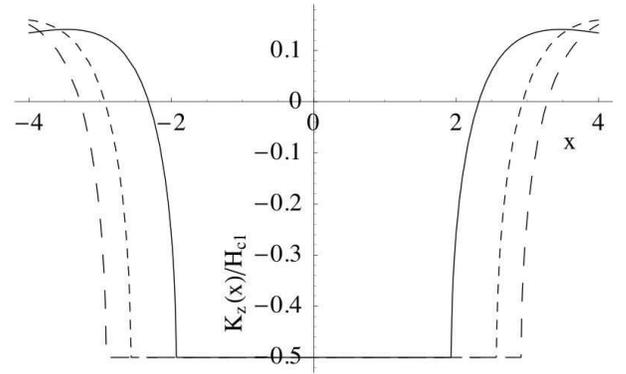}
\caption{%
Sheet-current density $K_z(x,0)/H_{c1}$ vs $x$ when $I_0 \ge 2I_{c1} +I_{c0}$ in
the presence of
bulk pinning,  shown here for $K_c=H_{c1}/2$,  $I_{c0} = 0.5 I_{c1}$,
$y_1 =1$ and
$y_2 = 2$. Shown are results for
$I_0 = 2.5 I_{c1}$ when
$a = 1.933$ (solid curve), 
$I_0 = 4.25 I_{c1}$ when $a = 2.504$ (short-dashed curve), and $I_0 = 6 I_{c1}$
when
$a = 2.913$ (long-dashed curve), corresponding to the cases shown in Fig.\ 
\ref{fig17}.}
\label{fig18}
\end{figure}

When $I_0 > 2I_{c1} + I_{c0}$, the gap
between the vortex dome and the antivortex dome is closed ($b = 0$), and the
magnetic-field distribution thus can be characterized as a dome of
vortices in the region $0 < x < a$ 
carrying magnetic flux up through the film and an adjacent dome of  antivortices
in the region $-a < x < 0$ carrying an equal amount of magnetic flux back down
through the film. The outer boundaries of these domes ($\pm a$) depend upon the
values of
$I_0$ and $K_c$, and the magnetic-field and sheet-current-density distributions
can be calculated as follows. 
The complex
magnetic field $\cH(\zeta) = \cH_M(\zeta) +\cH_v(\zeta)$ is as given in
Eq.~\eqref{cH_zero}, except that $b=0$ and  $\phi(\zeta)$, $s_1$, and
$s_2$ are now given by Eqs.~\eqref{phi0}, \eqref{s10}, and \eqref{s20}. 
Similarly, the vortex-generated complex magnetic field $\cH_v(\zeta)$ is as
given in Eq.~\eqref{Hvweak},  and 
$H_{vy}(x,0)$,
$H_{vx}(x,\epsilon)=-H_{vx}(x,-\epsilon),$ and
$K_{vz}(x)=H_{vx}(x,-\epsilon)-H_{vx}(x,\epsilon)$ are as given in
Eqs.~\eqref{Hvyx0_small}, \eqref{Hvxxeps_small}, and \eqref{Kzx_small},
except
that now just above or below the real axis,
\begin{equation}
	\phi(x \pm i\epsilon)= 
	\begin{cases} \displaystyle 
		\tilde \phi(x), & 0<|x|<a, \\
	\pm i	\tilde \phi(x),
		&  |x| > a, 
\end{cases}
\label{phixpm_Kcbig}
\end{equation}
where
\begin{equation}
	\tilde \phi(x)= 
	\begin{cases} \displaystyle 
	x \sqrt{a^2-x^2}, & 0<|x|<a, \\
-	|x|\sqrt{x^2-a^2}, & |x|>a .
	\end{cases}
\label{tildephi_Kcbig}
\end{equation}
For given values of $I_0$ and $K_c$ the value of $a$ in the above equations is
determined  by  Eq.~\eqref{condition3}, which follows from the requirement that
$\int_{-\infty}^{+\infty}K_z(x)dx = 0$, and by Eq.~\eqref{Hweakcondition} (but
with $b=0$), which follows from the requirement that
$H_x(0,\epsilon) = H_{c1}+K_c$.
Numerical solutions for $a$ obtained in this way are shown in
Fig.~\ref{fig11}  vs $I_0/I_{c1}$ for $I_0 > 2I_{c1} + I_{c0}$ (where $b=0$) for
the case of
$y_1 = 1$ and
$y_2 = 2$.
Figures \ref{fig17} and \ref{fig18} show plots of $H_y(x,0) = H_{vy}(x,0)$ and 
$K_z(x)$ vs
$x$ for several values of
$I_0 \ge 2I_{c1} + I_{c0}$ when $K_c = H_{c1}/2$.

When
$I_0 > 2I_{c1}+I_{c0}$, such that $a > 0$ and $b=0$, the vortex-generated complex
potential ${\cal G}_v(\zeta)={\cal G}(\zeta)-{\cal G}_M(\zeta)$ [see
Eqs.~\eqref{cG_linear}, \eqref{cG_zero}, and
\eqref{g0strong}] in the presence of bulk pinning  is  
\begin{eqnarray}
	{\cal G}_v(\zeta)\!\! &=\!\!& \frac{I_0}{2\pi} 
		\left[ g_v(\zeta,y_1) -g_v(\zeta,y_2) \right] 
 \pm i\frac{K_c}{2}\zeta , 
\label{cG_vstrong}
\end{eqnarray}
where 
\begin{eqnarray}
	g_v(\zeta,y) 
&	 =  \mp i\arctan (\zeta/y)+
	\sqrt{\frac{a^2-\zeta^2}{a^2+y^2}}-\frac{a}{\sqrt{a^2+y^2}} \nonumber \\
&
-\mbox{arctanh} \sqrt{\frac{a^2-\zeta^2}{a^2+y^2}}
+\mbox{arctanh} \frac{a}{\sqrt{a^2+y^2}},
\label{gvstrong}
\end{eqnarray}
and the
upper (lower) signs hold in the upper (lower) half $\zeta$ plane. 
\begin{figure}%***** Fig.19 ************************
\includegraphics[width=8cm]{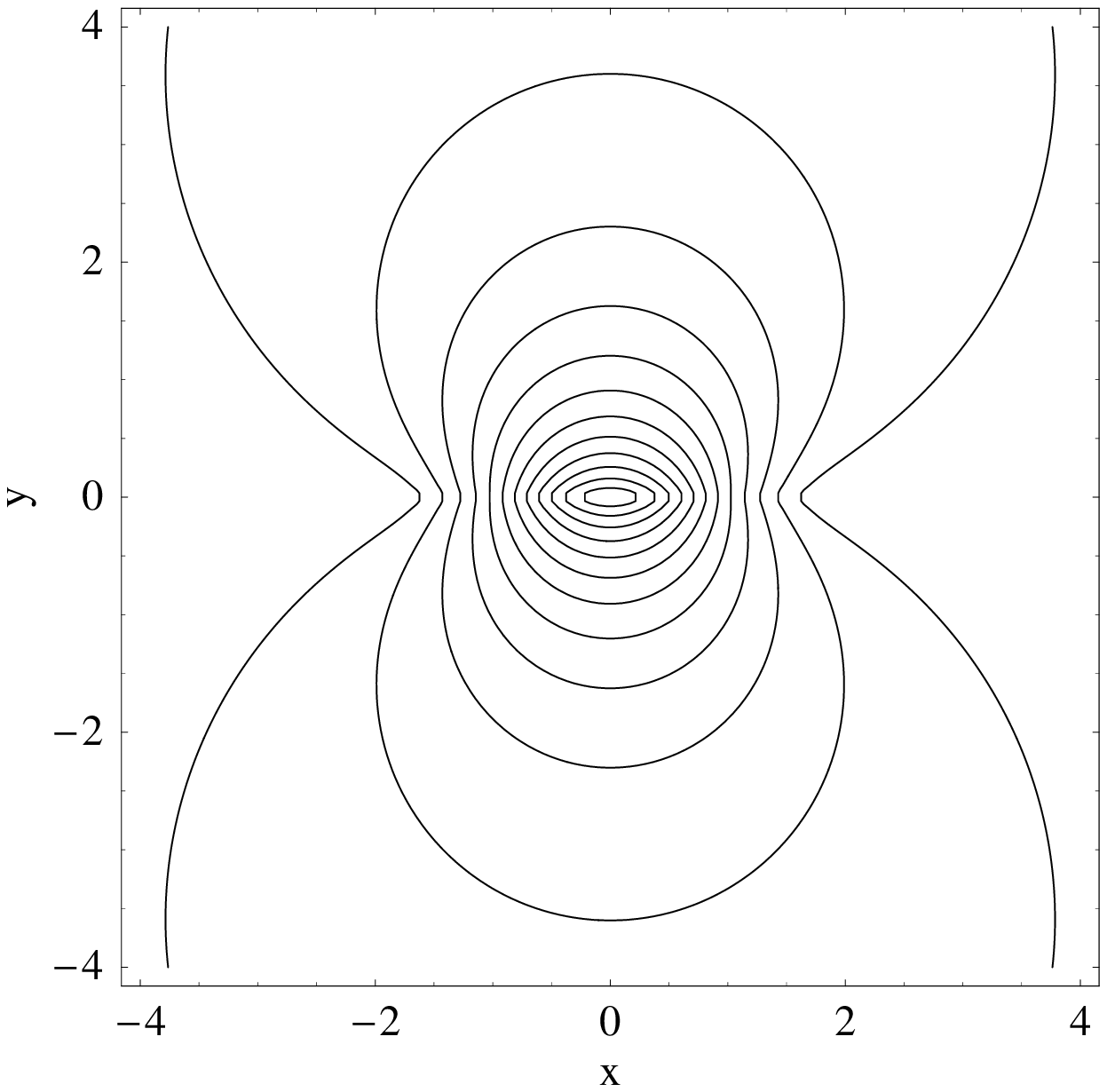}
\caption{%
Contour plot of the real part of the vortex-generated complex potential in the
presence of  bulk pinning [Eq.~\eqref{cG_vstrong}]
${\cal G}_v(x+iy)$ vs
$x$ and $y$ for $y_1 = 1$, $y_2 = 2$, $K_c = H_{c1}/2,$ $I_0 = 2 I_{c1}+I_{c0} =
2.5 I_{c1}$,
$a = 1.933$, and $b=0$, as for the solid curves in Figs.~\ref{fig17} and
\ref{fig18}.  The contours correspond to magnetic field lines of the
vortex-generated complex field
$\cH_v(x+iy)$ given by Eq.~\eqref{Hvweak}. }
\label{fig19}
\end{figure} 
Shown
in Fig.~\ref{fig19} is a contour plot of the real part of  ${\cal G}_v(x+iy)$. 
These contours correspond to the magnetic field lines of the
vortex-generated magnetic field.  The magnetic field flows in a
generally counterclockwise direction, carried by vortices  in the region $0 < x <
a$ up through the  film and by antivortices in the region $-a < x < 0$ back down
through the film.  A contour plot of the real part of  ${\cal G}(x+iy)=
{\cal G}_M(x+iy)+{\cal G}_v(x+iy)$ would also show the magnetic field lines
generated by the two wires, as in Fig.~2.

\subsection{Flux domes in the presence of strong bulk
pinning%**************************
\label{Sec_Kc_strong}} 

We consider here briefly the case of superconducting films in which bulk pinning,
characterized by a field-independent critical sheet-current
density 
$K_c  = j_c d >0$, is so strong that $K_c \gg H_{c1}$.  In principle, the process
of vortex entry is qualitatively the same as discussed in  Sec.~\ref{Sec_Kc>0}. 
However, in the limit that $H_{c1}/K_c \to 0$, the width of the current region
$I_{c1}+I_{c0}$ to 
$2I_{c1}+I_{c0}$ (in which there is a gap of width $2b$ between the vortex and
antivortex domes) shrinks to zero, and 
$I_{c0}$ becomes the only critical current of practical interest.  Essentially, as
soon as
$I_0$ exceeds $I_{c0}$, the  vortices penetrating from the top surface divide in
such a way as to produce adjacent vortex and antivortex flux domes in the regions
$0 < x < a$ and
$-a < x < 0$, and the sheet-current density is $K_z(x) = - K_c$ in these regions. 
The perpendicular magnetic field $H_y(x,0)$, sheet-current density $K_z(x)$ and
vortex-generated complex potential ${\cal G}_v(x+iy)$ can be calculated as
discussed in  Sec.~\ref{Sec_Kc>0} for the case that $b = 0$, and plots of all
these quantities look very similar to those in Figs.~\ref{fig17}, \ref{fig18},
and \ref{fig19}.

Figure 20 shows plots of
$H_x(0,\pm\epsilon)$ = $H_{Mx}(0,\pm\epsilon)+H_{vx}(0,\pm\epsilon)$ vs
$I_0$ in the limit $H_{c1}/K_c \to 0$.
\begin{figure}%***** Fig.20 ************************
\includegraphics[width=8cm]{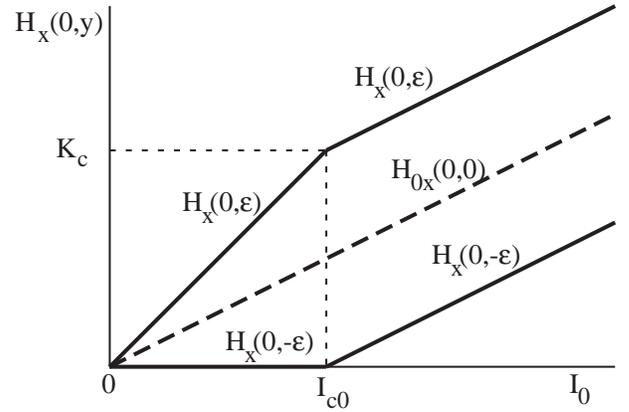}
\caption{%
Parallel magnetic fields at $x=0$ at the top and bottom
surfaces $H_x(0,\epsilon)$ and $H_x(0,-\epsilon)$ vs $I_0$ for 
strong bulk pinning
in the limit as $H_{c1}/K_c \to 0.$  When
$0
\le I_0
\le I_{c0}$,
$H_x(0,\epsilon) = H_{Mx}(0,\epsilon)$ [Eq.~\eqref{cH_linear}] and
$H_x(0,-\epsilon) = H_{Mx}(0,-\epsilon)=0$.  When $ I_0 \ge
I_{c0}$, $H_x(0,\pm\epsilon) = H_{0x}(0,0)\pm K_c/2$ [see Eq.~\eqref{H0}].}
\label{fig20}
\end{figure} 
\begin{figure}%***** Fig.21 ************************
\includegraphics[width=8cm]{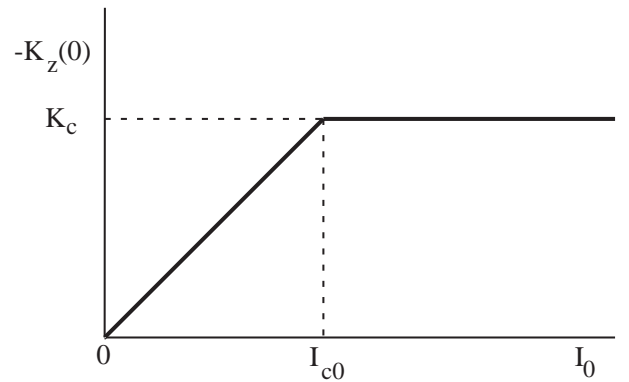}
\caption{%
Magnitude of the sheet-current density at the origin $-K_z(0) =
H_x(0,\epsilon)-H_x(0,-\epsilon)$ vs
$I_0$ for 
strong bulk pinning
in the limit as $H_{c1}/K_c \to 0$
 (see Fig.~20). }
\label{fig21}
\end{figure}
For $0 \le I_0 \le I_{c0}$, we have $H_x(0,\epsilon) = H_{Mx}(0,\epsilon) =
K_{c}I_0/I_{c0}$ and  $H_x(0,-\epsilon) = H_{Mx}(0,-\epsilon) =
0.$   For
$I_0 \ge I_{c0}$, $H_x(0,\pm \epsilon) = H_{0x}(0,0) \pm K_c/2 =
K_{c}I_0/2I_{c0} \pm K_c/2$.
Figure 21 shows a plot of
$-K_z(0)$ vs
$I_0$ in the same limit, where  $K_z(0) =
H_x(0,-\epsilon)-H_x(0,\epsilon)$.  For $0 \le I_0 \le I_{c0}$,
$-K_z(0) = K_{c}I_0/I_{c0}$, and for $I_0 \ge I_{c0}$, $-K_z(0)
= K_c$.

\section{Summary and Discussion%
\label{Sec_conclusion}}%************************** 

In this paper we have predicted that separated vortex and antivortex flux domes
can be produced in weak-pinning type-II superconducting films subjected to local
magnetic fields generated by current-carrying wires above the film's surface
and far from the edges.  To calculate these effects analytically, we have chosen
an idealized geometry of two parallel infinitely long wires above an infinite
superconducting film. However, the basic phenomenon of the creation of separated
vortex and antivortex flux domes without nucleation at the film's edges is far
more general than in the geometry we have considered here.

For example, consider the case of a bulk-pinning-free superconducting film of
finite size in the
$xz$ plane and a small coil at
$(x,y,z) = (0,y_0,0)$ a short distance above the film.  
Suppose the coil
produces a magnetic dipole moment
${\bm m} = -\hat x m$.  When $m$ is small, the film remains in the Meissner
state, and if the magnetic field below the film is very small, the coil-generated
dipole magnetic field and its image produce a magnetic field at the film's top
surface
${\bm H}_M =
\hat x H_{Mx} +
\hat z H_{Mz}$, where $H_{Mx}(x,\epsilon,z) = 2m(y_0^2-2x^2+z^2)/R_0^5$,
$H_{Mz}(x,\epsilon,z) = -6mxz/R_0^5$, and $R_0 = \sqrt{x^2+y_0^2+z^2}$.  The
induced sheet-current density is ${\bm K}_M = \hat x K_{Mx} +\hat z K_{Mz}$,
where $K_{Mx}(x,z) = H_{Mz}(x,\epsilon,z)$ and   $K_{Mz}(x,z) =
-H_{Mx}(x,\epsilon,z)$, such that $\nabla \cdot {\bm K}_M = 0$.
In the plane $z = 0$ the magnetic-field and sheet-current distributions resemble
those shown in Figs.~2 and 3.
The maximum magnetic field parallel to the top surface occurs at the origin, where
${\bm H}_M(0,\epsilon,0) = (2m/y_0^3)\hat x $, and the first flux
penetration through the film occurs when $m$ increases to the value $m_c =
H_{c1}y_0^3/2$.  The penetrating vortex splits into a vortex and an antivortex,
and the vortex is driven by the Meissner screening currents via the corresponding
Lorentz force
${\bm F}(x,z) = \phi_0 {\bm H}_M(x,\epsilon,z)$ to the point
$(x,y,z) = (y_0/\sqrt{2},0,0)$, while the antivortex is similarly driven to
$(x,y,z) = (-y_0/\sqrt{2},0,0)$.  A further increase in the coil's current ($m >
m_c$) will result in more nucleating vortices and
antivortices and cause the development of vortex and antivortex flux domes
centered at 
$(x,y,z) =
(\pm y_0/\sqrt{2},0,0)$.
Vortex and antivortex flux domes also could be produced by bringing a
permanent magnet with  magnetic dipole moment
${\bm m} = -\hat x m$ close to the film.

Similar effects should occur in superconducting films with bulk pinning. 
In particular, when the maximum magnetic field parallel to the surface,
accounting for image fields, exceeds $H_{c1} + K_c$, vortex and antivortex flux
domes should be produced with properties similar to those described in
Secs.~\ref{Sec_Kc>0} and
\ref{Sec_Kc_strong}.  In the above geometry with a small coil or a permanent
magnet producing a magnetic moment
${\bm m} = -\hat x m$, we expect that the main effect of bulk pinning will be to
reduce the separation between the vortex and antivortex flux domes.  

In this paper we have confined our attention to the case  in which an initially
flux-free film is subjected to locally applied magnetic fields increasing in
magnitude.  We expect that interesting hysteretic effects, similar to those  in a
finite-width film with a geometrical barrier,\cite{Benkraouda96}  will occur when 
ac magnetic fields are  applied.   For example, consider a bulk-pinning-free
superconducting film in the geometry studied in Sec.~\ref{Sec_SC-wire} but with 
$I_0$ being cycled.
As $I_0$ is increased from zero, we expect vortex and antivortex domes to develop
as predicted in Sec.~\ref{Sec_Kc=0}, where the dome boundaries
$a$ and $b$ are determined in part by the condition that $H_x(0,\epsilon) =
H_{c1}$ [Eq.~\eqref{I0ab}].  Suppose that these values are $a_{max}$ and $b_{max}$
when
$I_0$ increases to some maximum value $I_{max}$, where $I_{c1} < I_{max} < 2
I_{c1}$,
$a_{max} = x_0^2/b_{max} > x_0,$ and $b_{max} < x_0 = \sqrt{y_1 y_2}$.  If  $I_0$
is now decreased, the condition that
$H_x(0,\epsilon) = H_{c1}$ is replaced by the
condition that the magnetic flux under each flux dome remains constant; i.e.,  
$\Re[{\cal G}(a)-{\cal
G}(b)]$ = const.  As a result, as $I_0$ decreases, the width of each dome
increases while its height decreases; i.e., $b$ decreases ($b < b_{max}$) and
$a$ increases ($a > a_{max}$).  The resulting values of $b$ and $a$ can be
calculated as functions of $I_0$ using the relations $a = x_0^2/b$ and 
\begin{equation}
I_0 \Delta g(a,b) = I_{max} \Delta g(a_{max},b_{max}),
\end{equation}
where [see Eqs.~\eqref{cG_zero} and \eqref{g0}]
\begin{eqnarray}
\Delta g(a,b)& =& {\rm Re}\{[g_0(a,y_1)-g_0(a,y_2)] \nonumber \\
&&-[g_0(b,y_1)-g_0(b,y_2)]\}.
\end{eqnarray}
So long
as
$b >0$, vortex-antivortex annihilation cannot occur because the sheet
current flowing in the region $|x| < b$ still keeps vortices and antivortices
apart.  In fact, a subsequent increase of $I_0$ would produce reversible changes
in the width and height of the vortex and antivortex domes, provided $I_0 <
I_{max}$.  However, the vortex-free gap of width $2b$ closes when $I_0$ is
decreased to the value
$I_{ex}$ at which 
$b \to 0$ and
$a
\to
\infty$, where
\begin{equation}
I_{ex}=I_{max} \Delta g(a_{max},b_{max})/\Delta g(\infty,0)
\end{equation}
and $\Delta g(\infty,0) = \ln (y_2/y_1)$. 
When $I_0 = I_{ex}$,  the sheet-current density becomes everywhere zero, the film
appears as if it were completely incapable of screening, and the magnetic field
distribution (averaged over a length of the order of the intervortex
separation) is essentially the same as it would be in the absence of the film. 
On the other hand, viewing the field distribution as a linear superposition of the
Meissner-state response and a vortex-antivortex distribution, we see that when
the  gap of width $2b$ closes, vortex-antivortex annihilation begins to occur at
$x = 0$, and magnetic flux begins to exit from the vortex and antivortex domes.   
As
$I_0$ decreases from $I_{ex}$ to zero, the magnitude of the magnetic flux under
each dome  decreases to zero.
When  $I_0 = 0$, the film is again 
flux-free, and as
$I_0$  further  decreases to $-I_{max}$, the behavior is very  similar to that
for increasing $I_0$, except that the roles of vortices and
antivortices are interchanged.

We expect that similar but somewhat more complicated hysteretic effects will occur
in the presence of bulk pinning.  In the case of strong bulk pinning ($K_c \gg
H_{c1}$), the role of $H_{c1}$ can be neglected, and the hysteretic
properties can be calculated analytically as in
Ref.~\onlinecite{Mawatari06}, which treats the response of a superconducting film
to currents in linear wires in arrangements similar to that discussed in
Sec.~\ref{Sec_SC-wire}.  Such calculations illuminate the fundamental physics
underlying the ac technique
introduced by Claassen {\it et al.}\cite{Claassen91} to determine the critical
current density $j_c$ in superconducting films.  This technique employs a small 
coil, placed just above the film, carrying a sinusoidal current.  When the current
amplitude exceeds the value at which the maximum induced sheet-current
density reaches $K_c=j_c d$, a third-harmonic voltage appears in the
coil.\cite{Claassen91,Poulin93,Mawatari02,Yamasaki03} A similar technique was
introduced by Hochmuth and Lorenz.\cite{Hochmuth94}

The effects discussed in this paper and applied to type-II
superconducting films are quite general and also should  be observed in
type-I superconductors.  
Magnetic flux domes consisting of intermediate-state regions
containing multiply quantized flux tubes have been observed in type-I strips in
which the geometric barrier plays a dominant role.\cite{Castro99} 
It is therefore
likely that separated domes of positive and negative magnetic flux produced in
response to nearby current-carrying wires, coils, or permanent magnets also will 
be observed in weak-pinning type-I superconducting films, foils, or plates when
the net parallel field at the surface exceeds the bulk thermodynamic critical
field $H_c$ (or, when bulk pinning is present, $H_c +K_c$). In type-I
superconductors, however, we expect that the magnetic flux will enter in the form
of the intermediate state.  The analog of
a vortex dome will be an intermediate-state region consisting either of an array
of multiply quantized flux tubes or a meandering normal-superconducting domain
structure carrying magnetic flux up through the film, while the analog of an
antivortex dome will be a similar intermediate-state region carrying magnetic
flux down through the film.  Such magnetic structures should be observable by
magneto-optics or related means by placing the magnetic-field source on one side
of the sample and the magnetic-field detector on the opposite side.

\section*{ACKNOWLEDGMENTS} %***** 
We thank V. G. Kogan for stimulating
discussions.  This manuscript has been authored in part by Iowa State
University of Science  and Technology under Contract No.\ W-7405-ENG-82 with
the U.S.\ Department of  Energy.

\appendix %*****

\section{\label{A}%
Screening Effects} %******************
The primary purpose of this paper is to describe in detail how magnetic flux
penetrates through a superconducting film in the form of vortices and
antivortices, using the assumption that the magnetic field below the film is
initially negligibly small.  In the following we present results confirming that
this is an excellent approximation when $d > \lambda$ and $d \ll y_1$.

\begin{figure}%***** Fig.22 ************************
\includegraphics[width=8cm]{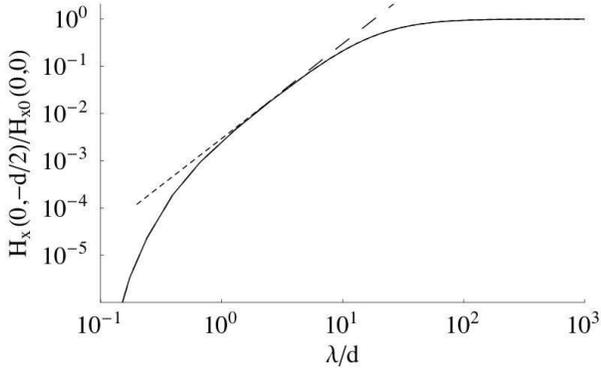}
\caption{%
Magnetic field just below the film $H_x(0,-d/2)/H_{x0}(0,0)$ vs $\lambda/d$,
shown here for $d = y_1/1000$ and $y_2 = 2y_1$ as calculated from Eq.~\eqref{Hbot}
(solid curve), Eq.~\eqref{Hbotsinh} (long-dashed curve), and
Eq.~\eqref{HbotLambda} (short-dashed curve). }
\label{fig22}
\end{figure}

\begin{figure}%***** Fig.23 ************************
\includegraphics[width=8cm]{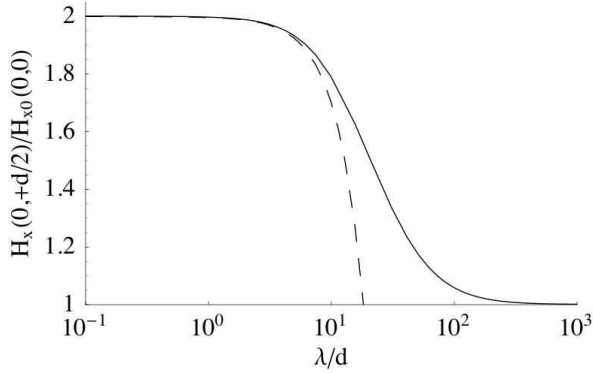}
\caption{%
Magnetic field just above the film $H_x(0,+d/2)/H_{x0}(0,0)$ vs $\lambda/d$,
shown here for $d = y_1/1000$ and $y_2 = 2y_1$ as calculated from
Eq.~\eqref{Htop} (solid curve) and Eq.~\eqref{Htoptanh} (dashed curve).  A plot of
the same quantity using Eq.~\eqref{HtopLambda} would be indistinguishable from the
solid curve.}
\label{fig23}
\end{figure}

Consider the experimental configuration shown in Fig.~1 but allow for the
finite thickness $d$ of the superconducting film in the region $|z| < d/2$.  When
the film, characterized by the London penetration depth $\lambda$, is in the
Meissner state, the vector potential and the magnetic field throughout all space,
as well as the supercurrent density in the film, can be calculated as described
in Ref.~\onlinecite{Clem92}.  The results for the $x$ components of the magnetic
fields
$H_x(0,\mp d/2)$ at the bottom and top surfaces of the film, expressed in units
of
$H_{0x}(0,0)$ [Eq.~\eqref{H0}], are
\begin{widetext}
\begin{eqnarray}
\frac{H_x(0,-d/2)}{H_{0x}(0,0)}&=&\frac{2 y_1 y_2}{(y_2 - y_1)}
\int_0^\infty \Big[ \frac{Qq}
{2qQ \cosh(Qd) +(q^2 + Q^2) \sinh(Qd)}\Big](e^{-qy_1}-e^{-qy_2})e^{qd/2} dq,
\label{Hbot} \\
\frac{H_x(0,+d/2)}{H_{0x}(0,0)}&=&\frac{2 y_1 y_2}{(y_2 - y_1)}
\int_0^\infty \Big[ \frac{Q[q \cosh(Qd)+Q \sinh(Qd)]}
{2qQ \cosh(Qd) +(q^2 + Q^2) \sinh(Qd)}\Big] (e^{-qy_1}-e^{-qy_2})e^{qd/2} dq,
\label{Htop}
\end{eqnarray} 
\end{widetext}
where $Q = \sqrt{q^2+\lambda^{-2}}$.  These quantities are plotted as the solid
curves in Figs.~\ref{fig22} and \ref{fig23} as functions of $\lambda/d$ for the
case that
$d = y_1/1000$ and $y_2 =2 y_1$.

When
$\lambda \to 0$, the film screens perfectly, such that
$H_x(0,-d/2)=0$ and 
$H_x(0,+d/2)=2H_{x0}(0,d/2)$.  In the opposite limit, when $\lambda \to \infty$,
$H_x(0,-d/2)=H_{x0}(0,-d/2)$ and $H_x(0,d/2)=H_{x0}(0,d/2)$, where 
$H_{x0}(0,y) = I_0(y_2-y_1)/2\pi( y_1-y)(y_2-y)$ is the magnetic field produced
by the wires  in the
plane
$x=0$ for
$y<y_1$ in the film's absence.

Equations \eqref{Hbot} and \eqref{Htop} reduce to simpler expressions
when $d \ll y_1$ and either $\lambda < d$ or, if $\lambda > d$, the Pearl
length\cite{Pearl64}
$\Lambda = \lambda^2/d$ obeys $\Lambda \ll y_1$.  Then to good approximation $q$
can be set equal to zero inside the brackets and $Q$ can be replaced by
$1/\lambda$.  The resulting integrals yield 
\begin{eqnarray}
\frac{H_x(0,-d/2)}{H_{0x}(0,0)}&=&\frac{2 (y_1 + y_2) \lambda}{y_1 y_2
\sinh{(d/\lambda)}}
\label{Hbotsinh} \\
\frac{H_x(0,+d/2)}{H_{0x}(0,0)}&=&2-\frac{2 (y_1 + y_2) \lambda}{y_1 y_2
\tanh{(d/\lambda)}}.
\label{Htoptanh}
\end{eqnarray} 
As shown by the long-dashed curves in Figs.~\ref{fig22} and
\ref{fig23}, these expressions are excellent approximations,
indistinguishable from the solid curves, when
$\lambda/d < 1$ or $\Lambda \ll y_1$ when $\lambda/d > 1$.  However,  the
long-dashed curves deviate significantly from the solid curves for $\lambda/d >
10$, which is to be expected, since for the parameters used for the figures,
$\Lambda \approx y_1$ when $\lambda/d \approx 30$.  

To evaluate Eqs.~\eqref{Hbot} and \eqref{Htop} for all values of the  Pearl
length\cite{Pearl64}   $\Lambda =
\lambda^2/d$
when
$d
\ll y_1$ and 
$\lambda \gg d$, it is a good approximation to ignore $q^2$ relative to
$1/\lambda^2$ inside the brackets and to replace
$Q$ by
$1/\lambda$ but to retain the terms proportional to $q$ in the denominators. 
This approximation is equivalent to the assertion that when  $d \ll \lambda$, the
only length that determines the screening properties of the film is $\Lambda$. 
This procedure yields the following approximate results:
\begin{eqnarray}
H_x(0,-d/2)/H_{0x}(0,0)&=&1-I,
\label{HbotLambda} \\
H_x(0,+d/2)/H_{0x}(0,0)&=&1+I,
\label{HtopLambda} 
\end{eqnarray}
where
\begin{eqnarray}
I&=&\frac{y_1 y_2}{(y_2 - y_1)}
\int_0^\infty\frac{1}
{1+2q\Lambda}(e^{-qy_1}-e^{-qy_2}) dq,
\label{F1} \\
&=&\frac{y_1 y_2}{2(y_2 - y_1)\Lambda}[e^{y_2/2\Lambda}{\rm{Ei}}(-y_2/2\Lambda)
\nonumber \\
&&\;\;\;\;\;\;\;\;\;\;\;\;\;\;\;\;\;\;\;
-e^{y_1/2\Lambda}{\rm{Ei}}(-y_1/2\Lambda)],
\label{F2}
\end{eqnarray} 
and Ei($x$) is the exponential integral.
On the scale of Fig.~\ref{fig22}, the approximation for $H_x(0,-d/2)$ given in
Eq.~\eqref{HbotLambda}, shown as the short-dashed curve, is
indistinguishable from the solid curve for values of $\lambda/d > 3$.
Surprisingly, on the
scale of Fig.~\ref{fig23}, the approximation for $H_x(0,d/2)$ given in
Eq.~\eqref{HtopLambda} is
indistinguishable from the solid curve for all values of 
$\lambda/d$.

\section{\label{B}%
Complex field and complex potential} %****************** 
In the following we derive the complex field $\cH(\zeta)$ and complex potential
$\cal G(\zeta)$ satisfying the boundary conditions that the perpendicular ($y$)
component of the magnetic field in the plane of the film obeys
$H_y(x,0)=0$ for
$|x| < b$ or
$|x| > a$ and that the sheet-current density in the $z$ direction obeys $K_z(x)
= - K_c$ for $b < |x| < a$.   We begin by considering the function
\begin{equation}
F(\zeta)=[\cH(\zeta)-\cH_{\|}]/\phi(\zeta),
\label{AF}
\end{equation}
where $\cH(\zeta)$ is defined in Eq.~\eqref{Biot-Savart},
\begin{equation}
\cH_{\|}(\zeta)=i	\frac{I_0}{2\pi}\Big(\frac{y_1}{\zeta^2+y_1^2}
-\frac{y_2}{\zeta^2+y_2^2}\Big),
\label{Hpara}
\end{equation}
and
\begin{equation}
\phi(\zeta)=\sqrt{(a^2-\zeta^2)(\zeta^2-b^2)}.
\label{Aphi}
\end{equation}
For $|\zeta| \to \infty$,
$\phi(\zeta) \to \mp i\zeta^2$, where the upper (lower) sign holds in the upper
(lower) half plane.  Just above (below) the real axis,
\begin{equation}
	\phi(x \pm i\epsilon)= 
	\begin{cases} \displaystyle 
	\pm i	\tilde \phi(x),
		& |x|<b  \mbox { or } |x| > a, \\
		\tilde \phi(x), & b<|x|<a, \\
	\end{cases}
\label{Aphixpm}
\end{equation}
where
\begin{equation}
	\tilde \phi(x)= 
	\begin{cases} \displaystyle 
	\sqrt{(a^2-x^2)(b^2-x^2)},
		& |x|<b , \\
		{\rm sgn}(x) \sqrt{(a^2-x^2)(x^2-b^2)}, & b<|x|<a, \\
-	\sqrt{(x^2-a^2)(x^2-b^2)}, & |x|>a .
	\end{cases}
\label{Atildephi}
\end{equation}
  As can be seen from Eqs.~\eqref{Biot-Savart} and
\eqref{Hpara}, $F(\zeta)$ is an analytic function of
$\zeta = x + iy$ except for poles at
$\zeta = \pm i y_1$ and $\zeta = \pm i y_2$, and a branch cut along the real
axis.

Next consider the following integral around the  closed contour $C$ consisting
of a line  just above the real axis at
$\zeta' = u + i \epsilon$ from $u=-\infty$ to $u=+\infty$, the infinite circle
at $\zeta' = R e^{i\theta}$ from $\theta = 0$ to $\theta = 2 \pi$ with $R\to
\infty$, and a line just below the real axis at $\zeta' = u - i \epsilon$ from
$u=+\infty$ to $u=-\infty$,
\begin{equation}
 \oint_C d\zeta' \frac{F(\zeta')}{\zeta'-\zeta}
=\int_{-\infty}^{+\infty}du\frac{F(u+i\epsilon)-F(u-i\epsilon)}{u-\zeta},
\label{Fintegral}
\end{equation}
where the integral around the infinite circle vanishes because $|F(\zeta)|
\propto |H(\zeta)/\zeta^2|\to 0$ as $|\zeta| \to \infty$.  Using the residue
theorem, accounting for the poles of the integrand on the left-hand side of
Eq.~\eqref{Fintegral} at $\zeta' = \zeta, i y_1, i y_2, -i y_1,$ and $-iy_2$, we
obtain
\begin{equation}
F(\zeta)-F_0(\zeta)
=\frac{1}{2 \pi i}
\int_{-\infty}^{+\infty}du\frac{F(u+i\epsilon)-F(u-i\epsilon)}{u-\zeta},
\label{F-F0}
\end{equation}
where 
\begin{eqnarray}
F_0(\zeta)&=&\frac{I_0}{2\pi}\Big(\frac{y_1}{s_1(\zeta^2+y_1^2)}
-\frac{y_2}{s_2(\zeta^2+y_2^2)}\Big),\\
	s_1&=&\sqrt{(a^2+y_1^2)(b^2+y_1^2)}, \\
	s_2&=&\sqrt{(a^2+y_2^2)(b^2+y_2^2)}.
\label{F0s1s2}
\end{eqnarray}
From Eqs.~\eqref{Biot-Savart}, \eqref{H0}, and \eqref{Hpara}, we find that 
\begin{equation}
\cH(x\pm i\epsilon)-\cH_{\|}(x\pm i\epsilon)=H_y(x,0)\mp i K_z(x)/2,
\end{equation}
such that Eqs.~\eqref{AF} and \eqref{Aphixpm} yield
\begin{equation}
F(u+i\epsilon)-F(u-i\epsilon)=
\begin{cases} \displaystyle 
	-2iH_y(x,0)/\tilde \phi(x),
		& |x|<b, \\
			-iK_z(x)/\tilde \phi(x), & b<|x|<a ,\\
	-2iH_y(x,0)/\tilde \phi(x), & |x|>a .\\
	\end{cases}
\label{Fdiff}
\end{equation}
However, $H_y(x,0) = 0$ for $|x|<b$ or $|x|>a$, and $K_z(x) = -K_c$ for $b <
|x| < a$.  Using Eqs.~\eqref{Atildephi} and \eqref{Fdiff} and evaluating the
integrals, we obtain the following expressions for the
complex field
$\cH(\zeta)$ and complex potential
${\cal G}(\zeta) =\int_{i\epsilon}^\zeta \cH(\zeta') d \zeta'$:
\begin{eqnarray}
	\cH(\zeta)\!\!&=\!\!&\frac{I_0}{2\pi} \Big\{ \frac{y_1[i
+\phi(\zeta)/s_1]}{\zeta^2+y_1^2} 
\! -\!\frac{y_2[i
+\phi(\zeta)/s_2]}{\zeta^2+y_2^2} 
		\Big\}\nonumber \\
&&\pm i K_c/2 ,
\label{cH_zero}\\
	{\cal G}(\zeta)\!\! &=\!\!& \frac{I_0}{2\pi} 
		\left[ g_0(\zeta,y_1) -g_0(\zeta,y_2) \right]
\pm i\frac{K_c}{2}\zeta , 
\label{cG_zero}
\end{eqnarray}
where 
\begin{eqnarray}
	g_0(\zeta,y) &=& \int_{\pm i\epsilon}^\zeta d\zeta' 
		\frac{y[i+\phi(\zeta')/s]}{\zeta'^2+y^2} 
\nonumber\\
	&=& i\arctan (\zeta/y) \nonumber\\
	&& {}\mp\frac{i}{asy} [ a^2y^2\bm{E}(\theta,k) 
		+y^2(b^2+y^2)\bm{F}(\theta,k)\nonumber\\
	&& -(a^2+y^2)(b^2+y^2)\bm{\Pi} 
		(\theta,-b^2/y^2,k) ],
\label{g0} \\
	s &=& \sqrt{(a^2+y^2)(b^2+y^2)}, \\
	\theta &=& \arcsin(\zeta/b), \mbox{ and } \\
	k &=& b/a , 
\end{eqnarray}
where $\bm E$, $\bm K$, and $\bm \Pi$ are incomplete elliptic integrals.

When $b=0$, the following replacements can be made in the above expressions,
\begin{eqnarray}
\phi(\zeta)&=&\zeta \sqrt{a^2-\zeta^2}, \label{phi0}\\
	s_1&=&y_1\sqrt{a^2+y_1^2},  \label{s10}\\
	s_2&=&y_2\sqrt{a^2+y_2^2}, \label{s20}
\end{eqnarray}
and  $\bm E$, $\bm K$, and $\bm \Pi$ can be evaluated to obtain
\begin{eqnarray}
	g_0(\zeta,y) 
&	 =  i\arctan (\zeta/y)+
	\sqrt{\frac{a^2-\zeta^2}{a^2+y^2}}-\frac{a}{\sqrt{a^2+y^2}} \nonumber \\
&
-\mbox{arctanh} \sqrt{\frac{a^2-\zeta^2}{a^2+y^2}}
+\mbox{arctanh} \frac{a}{\sqrt{a^2+y^2}}.
\label{g0strong}
\end{eqnarray}

It follows from Eq.~\eqref{Biot-Savart} that the 
integral $\int \cH(\zeta)
d\zeta$ around the great circle at $|\zeta| \to \infty$ yields $i \int K_z(x)
dx$ along the real axis.  The requirement that the film carries no
net current is thus equivalent to the requirement that
$\int \cH(\zeta)
d\zeta = 0.$  Using Eq.~\eqref{cH_zero} and the property that for
$|\zeta| \to
\infty$,
$\phi(\zeta) \to \mp i\zeta^2$, where the upper (lower) sign holds in the upper
(lower) half plane, we obtain the
requirement that when $K_c = 0$, 
$y_1/s_1 =y_2/s_2$.  Solving the latter equation, we obtain the following
condition relating $a$ and $b$,
\begin{equation}
ab = x_0^2=y_1 y_2.
\label{condition1}
\end{equation}
Similarly, when $ K_c>0$ and
$b >0$, the requirement that the film carries no net current leads to the
condition that 
\begin{equation}
\frac{I_0}{\pi}\Big(\frac{y_1}{\sqrt{(a^2+y_1^2)(b^2+y_1^2)}}
-\frac{y_2}{\sqrt{(a^2+y_2^2)(b^2+y_2^2)}}\Big)=K_c.
\label{condition2}
\end{equation}
When $ K_c>0$ but
$b =0$, the same requirement leads to the condition that 
\begin{equation}
\frac{I_0}{\pi}\Big(\frac{1}{\sqrt{a^2+y_1^2}}-\frac{1}{\sqrt{a^2+y_2^2}}\Big)=K_c.
\label{condition3}
\end{equation}

\end{document}